# Three Warm Jupiters orbiting TOI-6628, TOI-3837, TOI-5027 and one sub-Saturn orbiting TOI-2328⋆

Marcelo Tala Pinto[1,2], Andrés Jordán[1,2,3,4], Lorena Acuña[5], Matías Jones[6], Rafael Brahm[1,2,3], Yared Reinarz[5], Jan Eberhardt[5], Néstor Espinoza[7], Thomas Henning[5], Melissa Hobson[8,5,2], Felipe Rojas[9,2], Martin Schlecker[10], Trifon Trifonov[5,11,12], Gaspar Bakos[13], Gavin Boyle[14,4], Zoltan Csubry[13], Joel Hartmann[13], Benjamin Knepper[15,2], Laura Kreidberg[5], Vincent Suc[1,4], Johanna Teske[16,17], R. Paul Butler[16], Jeffrey Crane[17], Steve Schectman[17], Ian Thompson[17], Dave Osip[18], George Ricker[19], Karen A. Collins[20], Cristilyn N. Watkins[20], Allyson Bieryla[20], Chris Stockdale[21], Gavin Wang[22], Roberto Zambelli[23], Sara Seager[19,24,25], Joshua Winn[13], Mark E. Rose[26], Malena Rice[27], and Zahra Essack[28]

[1] Facultad de Ingeniería y Ciencias, Universidad Adolfo Ibáñez, Av. Diagonal las Torres 2640, Peñalolén, Santiago, Chile
e-mail: marcelo.tala@edu.uai.cl
[2] Millennium Institute for Astrophysics, Santiago, Chile
[3] Data Observatory Foundation, Santiago, Chile
[4] El Sauce Observatory — Obstech, Coquimbo, Chile
[5] Max-Planck-Institut für Astronomie, Königstuhl 17, 69117 Heidelberg, Germany
[6] European Southern Observatory (ESO), Avenida Alonso de Córdova 3107, Vitacura, Santiago, Chile
[7] Space Telescope Science Institute, 3700 San Martin Drive, Baltimore, MD 21218, USA
[8] Observatoire de Genève, Département d'Astronomie, Université de Genève, Chemin Pegasi 51b, 1290 Versoix, Switzerland
[9] Instituto de Astrofísica, Facultad de Física, Pontificia Universidad Católica de Chile, Av. Vicuña Mackenna 4860, Santiago, Chile.
[10] Department of Astronomy/Steward Observatory, The University of Arizona, 933 North Cherry Avenue, Tucson, AZ 85721, USA
[11] Landessternwarte, Zentrum für Astronomie der Universtät Heidelberg, Königstuhl 12, 69117 Heidelberg, Germany
[12] Department of Astronomy, Sofia University St Kliment Ohridski, 5 James Bourchier Blvd, BG-1164 Sofia, Bulgaria
[13] Department of Astrophysical Sciences, Princeton University, Princeton, NJ 08544, USA
[14] Cavendish Laboratory, J J Thomson Avenue, Cambridge, CB3 0HE, UK
[15] Department of Physics, Cornell University, Ithaca, New York, U.S.A.
[16] Carnegie Institution for Science, Earth & Planets Laboratory, 5241 Broad Branch Road NW, Washington DC 20015, USA
[17] The Observatories of the Carnegie Institution for Science, 813 Santa Barbara Street, Pasadena, CA 91101
[18] Las Campanas Observatory, Carnegie Institution of Washington, Colina el Pino, Casilla 601 La Serena, Chile
[19] Department of Physics and Kavli Institute for Astrophysics and Space Research, Massachusetts Institute of Technology, 77 Massachusetts Avenue, Cambridge, MA 02139, USA
[20] Center for Astrophysics | Harvard & Smithsonian, 60 Garden Street, Cambridge, MA 02138, USA
[21] Hazelwood Observatory, Australia
[22] Department of Physics & Astronomy, Johns Hopkins University, 3400 N. Charles Street, Baltimore, MD 21218, USA
[23] Società Astronomica Lunae, Castelnuovo Magra, Italy
[24] Department of Earth, Atmospheric and Planetary Sciences, Massachusetts Institute of Technology, 77 Massachusetts Avenue, Cambridge, MA 02139, USA
[25] Department of Aeronautics and Astronautics, Massachusetts Institute of Technology, 77 Massachusetts Avenue, Cambridge, MA 02139, USA
[26] NASA Ames Research Center, Moffett Field, CA 94035, USA
[27] Department of Astronomy, Yale University, 219 Prospect Street, New Haven, CT 06511, USA
[28] Department of Physics and Astronomy, The University of New Mexico, 210 Yale Blvd NE, Albuquerque, NM 87106, USA

## ABSTRACT

We report the discovery and characterization of three new transiting giant planets orbiting TOI-6628, TOI-3837 and TOI-5027, and one new warm sub-Saturn orbiting TOI-2328, whose transits events were detected in the lightcurves of the Transiting Exoplanet Survey Satellite space mission. By combining TESS lightcurves with ground-based photometric and spectroscopic follow-up observations we confirm the planetary nature of the observed transits and radial velocity variations. TOI-6628 *b* has a mass of 0.75±0.06 $M_J$ and is orbiting a metal-rich star with a period of 18.18424±0.00001 days and an eccentricity of 0.667±0.016, making it one of the most eccentric orbits of all known warm giants. TOI-3837 *b* has a mass of 0.59±0.06 $M_J$ and orbits its host star every 11.88865±0.00003 days, with a moderate eccentricity of $0.198^{+0.046}_{-0.058}$. With a mass of 2.01±0.13 $M_J$, TOI-5027 *b* orbits its host star in an eccentric orbit with $e = 0.395\pm^{0.032}_{0.029}$ every 10.24368±0.00001 days. TOI-2328 *b* is a Saturn-like planet with a mass of 0.16±0.02 $M_J$ orbiting its host star in a nearly circular orbit with $e = 0.057\pm^{0.046}_{0.029}$ at an orbital period of 17.10197±0.00001 days. All four planets have orbital periods above 10 days, and our planet interior structure models are consistsent a rocky-icy core with a H/He envelope, providing evidence supporting the core accretion model of planet formation for this kind of planets.

## 1. Introduction

The increasing number and diversity of exoplanets discovered in the last decades has raised fundamental questions about their formation and evolution mechanisms.

Many of these planets are giant planets orbiting very close to their host star, despite the low occurrence rate of such population (Stevenson 1982; Mayor et al. 2011; Fressin et al. 2013; Santerne, A. et al. 2016). This is mainly driven by the inherent observational bias in their detection, given their small periods, which results in a dominant population of Hot Jupiters (HJs) - highly irradiated planets given their proximity to their host star - which constitute about 75% of the well characterized transiting planets - planets with radii and masses measured with an uncertainty better than about 20%. Most of such discoveries were made by wide-field ground-based observatories (Bakos et al. (2006), Mandushev et al. (2005), Udalski et al. (2002)), by the Kepler and K2 space missions (Borucki (2017), Howell et al. (2014)) and, more recently, by the Transiting Exoplanet Survey Satellite mission (TESS, Ricker et al. (2015)).

Giant planets are thought to form via two different mechanisms: either by accreting its mass from the proto-planetary disk (Pollack et al. 1996), or by gravitational instabilities, in which the proto-planetary disk fragments into bound clumps to form a planet (Cameron 1978). Both mechanisms predict a low formation rate at close-in orbits. Close to the star, gas conditions could prevent the formation of bound clumps (Rafikov 2005), while at the same time such conditions may prevent the formation of sufficiently large cores to accrete gas from the proto-planetary disk (Schlichting 2014).

Therefore, the observed population of HJs strongly suggests that they might have formed further out, in regions where the conditions for core accretion and gravitational instabilities are more favorable, and migrated inwards into their currently close-in circular orbits.

Possible migration mechanisms consider interactions with the disc (e.g. Walsh et al. (2011)) and tidal migration produced by changes in the orbit, which can be induced by planet-planet scattering (Rasio & Ford 1996) or cyclic and/or chaotic secular interactions (Kozai 1962; Lidov 1962; Wu & Lithwick 2011).

Both mechanisms predict different orbital parameters for the planet population. While disc migration predict mainly circular orbits aligned with the stellar spin, high eccentricity migration predicts a wide distribution of eccentriticies and obliquities (Chatterjee et al. 2008). In this context, the estimation of the bulk structures and atmospheric parameters of such populations of exoplanets could provide crucial information about their formation locations in the disc and possible migration paths (Madhusudhan et al. 2014; Mollière et al. 2022).

However, the characterization of the formation history of HJs, widely defined as planets with periods smaller than 10 days, can be problematic, as their proximity to their host star may induce radiative and tidal interactions which can affect the interpretation of the origins of the observed population (e.g. Albrecht et al. 2012). Therefore, giant planets within the snow-line, but at farther distances from their host star, represent a unique opportunity to study possible formation and migration mechanisms in giant planets, as they are significantly less affected by their proximity to their host star, allowing a more direct comparison of their orbital properties with formation and migration models.

Warm Jupiters (WJs), broadly defined as giants planets orbiting their host star with periods longer than 10 days, remained elusive to ground-based transit searches, mainly due to the constrains imposed by the daily cycle. While some detections were possible with the Kepler+K2 missions (i.e. Huang et al. (2016)), only with TESS it has been possible to systematically characterize them, either through the detection of periodic transiting signals or single transiters (i.e. Brahm et al. (2019a), Gill et al. (2020)). By complementing transit observations with radial velocity (RV) measurements, it is possible to provide a detailed characterization of the dynamical and physical properties of the planetary system.

The Warm gIaNts with TESS (WINE) collaboration is a dedicated survey to identify, confirm and characterize WJs using TESS data and ground-based photometric and spectroscopic follow-up facilities, with the main goal of building a WJs database to constrain theories of planetary formation and evolution (Kossakowski et al. 2019; Jordán et al. 2020; Brahm et al. 2020; Schlecker et al. 2020; Hobson et al. 2021; Trifonov et al. 2021, 2023; Brahm et al. 2023; Hobson et al. 2023; Jones et al. 2024).

Here we present the discovery, confirmation, and orbital characterization of four new transiting exoplanets by the WINE collaboration, three of them WJs and one Saturn-like planet.

The paper is organized as follows. In Sect. 2, we present our observations. In Sect. 3, we present stellar parameters of the stars under study, in Sect. 4 the orbital parameters of the planetary systems and in Sect. 5 the modelling of the planetary interiors. We provide a discussion of our results in Sect. 6 and in Sect. 7 we present our conclusions.

## 2. Observations

### 2.1. TESS

All four transiting candidates presented in this study were identified from the full-frame image (FFI) light curves of the TESS primary mission. The FFIs were calibrated by the Science Processing Operations Center (SPOC) at NASA Ames Research Center (Jenkins et al. 2016).

TOIs 6628, 3837 and 5027 were also detected by the FAINT search pipeline (Kunimoto & Daylan 2021) and alerted to the community by the TESS Science Office on 15 August 2023, 23 June 2021, and 6 January 2022 after review by the TESS Science Office (Guerrero et al. 2021). TOI-2328 was also detected by the SPOC transit search pipeline and alerted to the community on 7 October 2020. The difference image centroiding analysis (Twicken et al. 2018) presented in the SPOC Data Validation reports for TOIs 3837, 5027 and 2328 constrained the location of the source of the transit signatures to within 0.219±2.4 arcsec (Sector 45), 0.951±2.5 arcsec (Sector 66), and 0.330±2.5 arcsec (Sectors 1-69), respectively.

Due to our focus on the detection of giant transiting giant planets in long period orbits (P>10 days), we process all the light curves with a dedicated algorithm that, after passing them through a median filter, identifies systematic negative deviations in flux, whose amplitudes could be consistsent with the depths and durations of transits of giant planets orbiting main-sequence/subgiant stars. All systems that present such specific variations are then manually vetted. The possible dilution of the transits by neighboring stars that fall inside the TESS photometric aperture is considered in the manual vetting process when estimating the planet radius from the transit depth and host stellar radius.

⋆ Based on observations collected at La Silla Observatory under programs IDs 0103.A-9008(A), 0104.A-9007(A), 0104.C-0413(A), 0105.A-9001(A), 0106.A-9014(A), 0109.A-9003(A), 0110.A-9011(A), 0111.A-9011(A), 105.20GX.001, 106.21ER.001, 108.22A8.001 and 109.239V.001, and through the Chilean Telescope Time under programmes CN2020B-21, CN2021A-14 and CN2021B-23. This paper includes data gathered with the 6.5 meter Magellan Telescopes located at Las Campanas Observatory, Chile.x

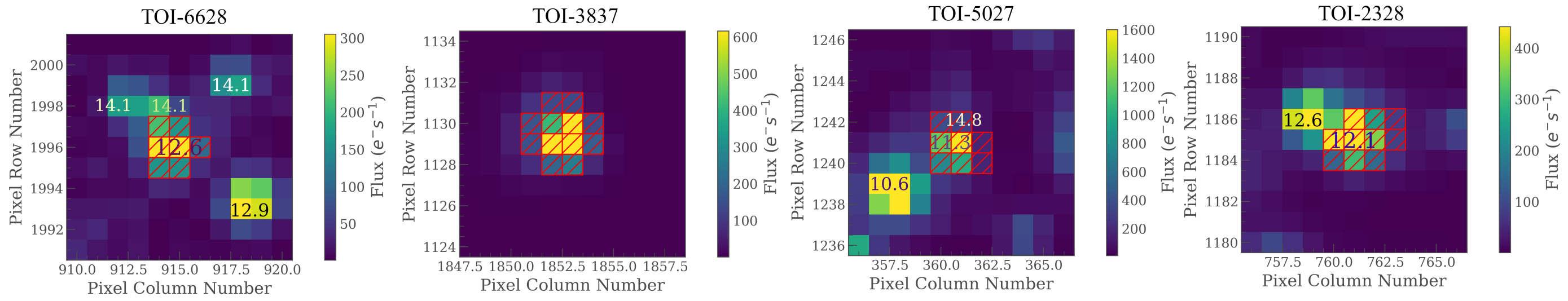

**Fig. 1.** TESS target pixel files (TPF) of TOI-6628, TOI-3837, TOI-5027 and TOI-2328. We have shown in red striped squares the aperture mask used in the lightcurve measurement. We have indicated the Gaia (Gaia Collaboration et al. 2018) magnitude of the surrounding stars.

| Parameter | TOI-6628 | TOI-3837 | TOI-5027 | TOI-2328 |
|---|---|---|---|---|
| TIC ID[1] | 75650448 | 308073360 | 361711730 | 394287035 |
| RA (J2015.5)[2] | 15h03m20.89s | 10h10m39.91s | 16h15m32.41s | 02h06m45.78s |
| Dec (J2015.5)[2] | -35d13m50.08s | +24d16m23.24s | -69d13m01.55s | -81d14m50.55s |
| $pm_{RA}$ (mas/yr)[2] | -30.17±0.09 | 50.37±0.07 | -22.02±0.03 | 31.80±0.03 |
| $pm_{Dec}$ (mas/yr)[2] | -15.94±0.06 | -37.02±0.07 | -31.99±0.04 | -1.67±0.03 |
| $\pi$ (mas)[2] | 3.09±0.06 | 3.18±0.04 | 4.86±0.02 | 4.28±0.02 |
| T (mag)[1] | 12.084±0.006 | 11.366±0.007 | 10.910±0.006 | 11.584±0.006 |
| V (mag)[3] | 12.771±0.068 | 11.960±0.028 | 11.512±0.101 | 12.297±0.008 |
| B (mag)[3] | 13.801±0.043 | 12.578±0.097 | 12.083±0.024 | 13.096±0.011 |
| G (mag)[2] | 12.640±0.002 | 11.804±0.002 | 11.334±0.002 | 12.108±0.002 |
| J (mag)[4] | 11.317±0.025 | 10.715±0.023 | 10.316±0.026 | 10.842±0.023 |
| H (mag)[4] | 10.943±0.026 | 10.401±0.021 | 10.042±0.023 | 10.458±0.021 |
| K (mag)[4] | 10.836±0.023 | 10.316±0.017 | 9.979±0.023 | 10.395±0.021 |
| Distance (pc)[5] | 315.2±8.3 | 306.7±5.8 | 202.2±1.5 | 228.9±1.5 |
| Luminosity ($L_\odot$)[5] | $0.817^{+0.046}_{-0.044}$ | $1.571^{+0.085}_{-0.076}$ | $0.932^{+0.034}_{-0.033}$ | $0.676^{+0.031}_{-0.031}$ |
| Mass ($M_\odot$)[5] | $0.97^{+0.06}_{-0.06}$ | $0.89^{+0.05}_{-0.05}$ | $1.02^{+0.05}_{-0.05}$ | $0.95^{+0.06}_{-0.06}$ |
| Radius ($R_\odot$)[5] | $1.01^{+0.06}_{-0.05}$ | $1.20^{+0.05}_{-0.05}$ | $0.92^{+0.04}_{-0.04}$ | $0.90^{+0.06}_{-0.06}$ |
| $\rho_*$ (g/cm$^3$)[5] | $1.323\pm^{0.225}_{0.187}$ | $0.725\pm^{0.118}_{0.098}$ | $1.831\pm^{0.038}_{0.044}$ | $1.828\pm^{0.289}_{0.245}$ |
| $T_{\rm eff}$ (K)[5] | $5463^{+143}_{-142}$ | $5905^{+157}_{-155}$ | $5909^{+145}_{-144}$ | $5525^{+147}_{-144}$ |
| Age (Gyr)[5] | $7.5^{+2.4}_{-2.5}$ | $11.2^{+2.5}_{-2.6}$ | $0.4^{+0.6}_{-0.3}$ | $4.0^{+2.3}_{-2.7}$ |
| $\log g$ (cm s$^{-2}$)[5] | $4.416^{+0.024}_{-0.025}$ | $4.229^{+0.018}_{-0.019}$ | $4.517^{+0.006}_{-0.008}$ | $4.505^{+0.021}_{-0.020}$ |
| $A_v$ (mag) | $0.313^{+0.083}_{-0.084}$ | $0.228^{+0.072}_{-0.068}$ | $0.100^{+0.041}_{-0.041}$ | $0.262^{+0.063}_{-0.060}$ |
| [Fe/H] (dex)[5] | 0.36±0.02 | -0.26±0.05 | -0.15±0.03 | 0.09±0.03 |
| $v \sin i$ (km s$^{-1}$)[5] | 3.13±0.20 | 3.47±0.26 | 4.52±0.13 | 2.61±0.25 |

**Table 1.** Stellar parameters. References: 1. TESS (Ricker et al. 2015), 2. Gaia Collaboration et al. (2018), 3. APASS (Munari et al. 2014), 4. 2MASS (Skrutskie et al. 2006), 5. This work. Uncertainties refer to the 68% credibility interval.

We also obtained 2-minute cadence light curves from the Mikulski Archive for Space Telescopes (MAST). We use the data calculated by the SPOC, which provides simple aperture photometry (SAP) and systematics-corrected Presearch Data Conditioning photometry (PDC, Smith et al. (2012), Stumpe et al. (2012), Stumpe et al. (2014)). We also use data calculated by the *Quick Look Pipeline* (QLP, Huang et al. (2020a,b); Kunimoto et al. (2021, 2022)).

To study possible contamination from other sources, we analyzed the Target Pixel Files (TPF). Fig. 1 shows the TPF for the sectors in which the first transit was detected for the four targets under study. The TPF plots show the field around the observed target, and the aperture mask used in the lightcurves construction. The PDC-SAP data from MAST has already been corrected by the contamination of neighboring stars and instrumental systematics, therefore no dilution correction is applied to these lightcurves.

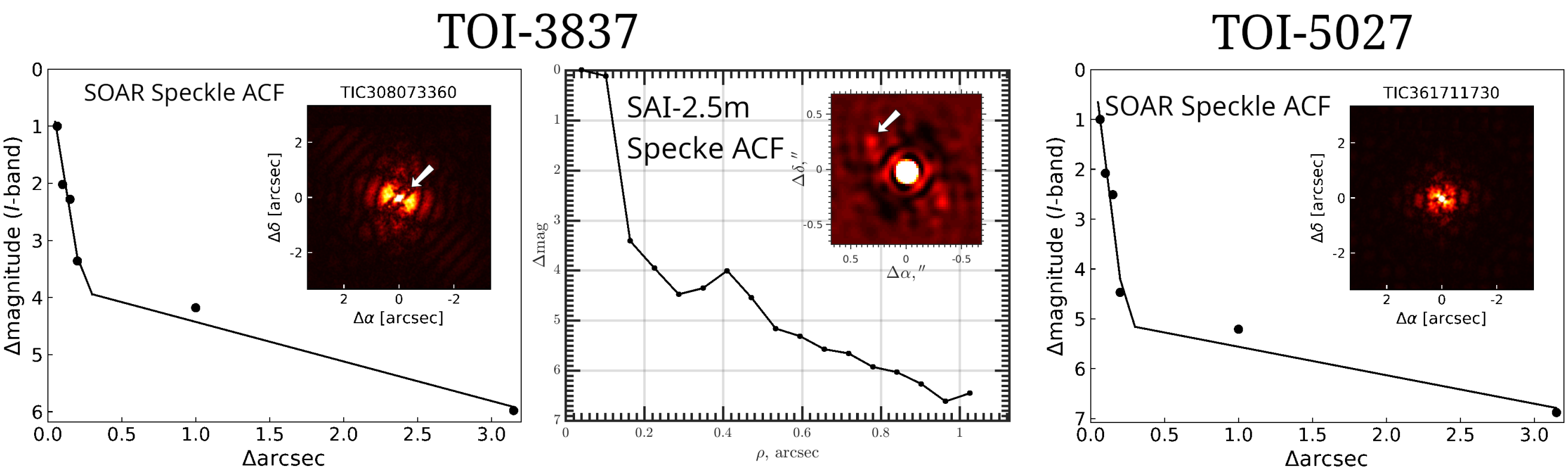

**Fig. 2.** High resolution images of TOI-3837 and TOI-5027.

| Observatory | Aperture (m) | Location | UTC Date | Filter | Transit coverage | Planet |
|---|---|---|---|---|---|---|
| OM-ES CDK600 | 0.6 | Observatorio El Sauce | 2021 June 19th | $r'$ | Full transit | TOI-6628 *b* |
| OM-ES RiDK500 | 0.5 | Observatorio El Sauce | 2024 March 15th | $r'$ | Full transit | TOI-6628 *b* |
| Hazelwood | 0.5 | Hazelwood Observatory | 2024 April 20th | Rc | Ingress | TOI-6628 *b* |
| LCOGT | 0.35 | CTIO | 2024 May 27th | ip | Transit and egress | TOI-6628 *b* |
| KeplerCam | 1.2 | FLWO | 2022 April 27th | $i'$ | Egress | TOI-3837 *b* |
| LCOGT | 1.0 | CTIO | 2022 April 3rd | $i'$ | Transit and egress | TOI-5027 *b* |
| OM-ES RiDK500 | 0.5 | Observatorio El Sauce | 2024 March 21 | $r'$ | Full transit | TOI-5027 *b* |
| CHAT | 0.7 | LCO | 2019 December 26th | $r'$ | Egress | TOI-2328 *b* |
| LCOGT | 1.0 | SSO | 2020 September 24th | $i'$ | Full transit | TOI-2328 *b* |
| Hazelwood | 0.32 | Hazelwood Observatory | 2021 February 8th | Rc | Egress | TOI-2328 *b* |
| LCOGT | 1.0 | SSO | 2021 August 15th | $i'$ | Egress | TOI-2328 *b* |
| LCOGT | 1.0 | SSO | 2021 August 15th | $g'$ | Egress | TOI-2328 *b* |
| OM-ES CDK600 | 0.6 | El Sauce Observatory | 2021 October 23rd | $r'$ | Egress | TOI-2328 *b* |

**Table 2.** Details of the ground-based follow-up observations obtained for the targets presented here.

All four targets have been selected as a high-priority candidate of the WINE collaboration, suitable for spectroscopic follow-up.

TOI-6628 was observed by TESS in sectors 11, 38 and 65. We have identified five transit events with an average depth of 7500 ppm and a duration of ~0.2 days in the TESS lightcurves. A query to the Gaia DR2 archive (Gaia Collaboration et al. 2018) revealed five neighboring stars within a 15” radius with Gaia magnitudes larger than 18. The brightest star close to TOI-6628 has a magnitude of 12.91. In addition, we observe three other stars in the nearby field with magnitudes of about 14 mag. Such targets can be observed in the TESS target pixel files, shown in Fig. 1, together with the aperture mask employed by the TESS pipeline, shown in red striped squares. Given the proximity of the 14.13 mag star to our target, we may expect some minor contamination from this target in one of the pixels of the aperture mask. We acknowledge, though, that TOI-6628 is also observed from ground-based telescopes with a much larger pixel scale, avoiding contamination from nearby objects, confirming the transiting signal in TOI-6628.

TOI-3837 was observed by TESS in sector 21, and then revisited in sectors 45, 46 and 48. We have identified seven transits in the TESS lightcurves, with an average depth of about 7000 ppm and a transit duration of 0.22 days. A query to the Gaia DR2 archive reveals no neighbor star within a 30” radius, which is in agreement with the TESS TPF image shown in Fig. 1.

TOI-5027 was observed by TESS in sector 12, 39 and 66. Six transit events with an average depth of about 10800 ppm and a duration of ~0.14 days was identified in TESS lightcurves. A query to the Gaia DR2 archive revealed six neighboring stars closer than 30”. All such neighboring targets have magnitudes larger than 18. We do observe a nearby bright companion with a Gaia magnitude of 10.61, but it is located outside the TESS aperture masks.

TOI-2328 was observed by TESS in sector 1, 13, 27, 28, 39, 67 and 68. We have identified ten transits in the TESS lightcurves, with a transit depth of about 10400 ppm and a transit duration of about 0.15 days. A query to the Gaia data archive reveals one neighbor star within a 15” radius, with a magnitude of about 17. An additional bright target with a Gaia magnitude of 12.62 is located at about 1 arcmin distance from our target, but it does not contaminate the aperture mask. This target can be observed in the sector 13 TPF image, shown in Fig. 1.

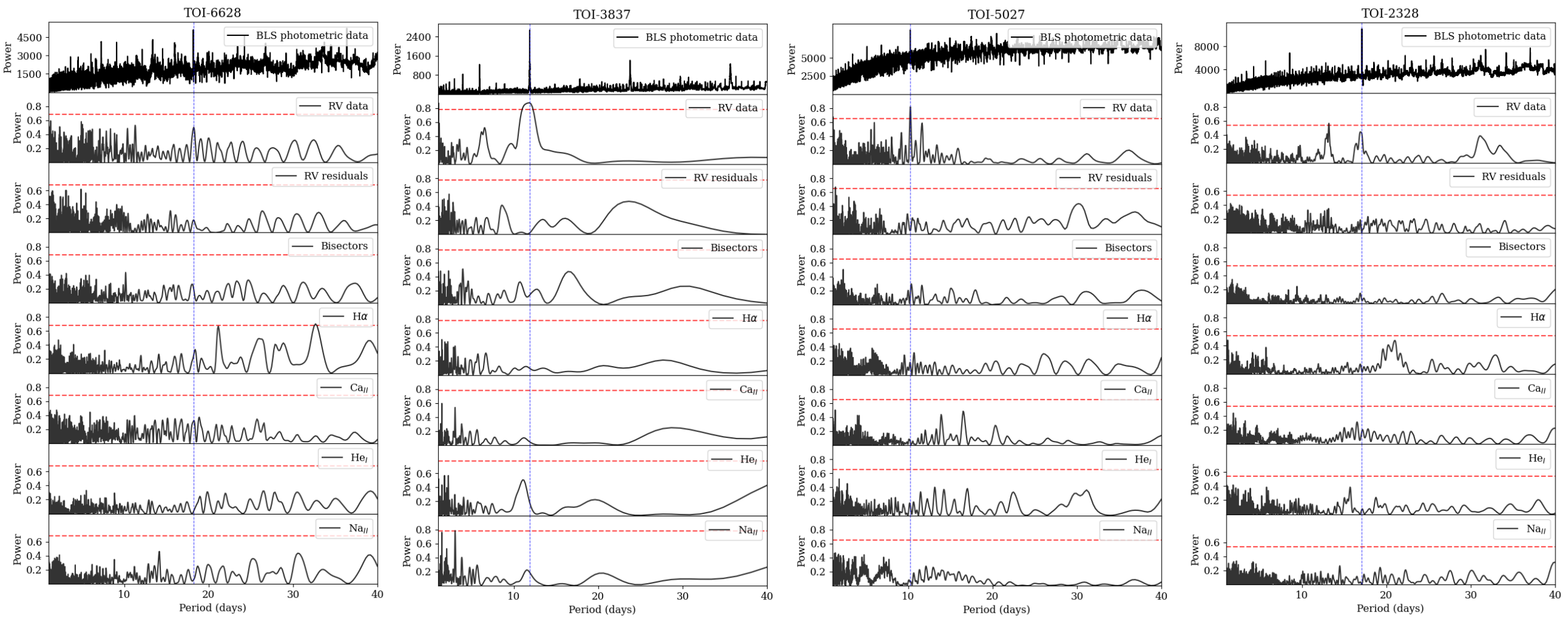

**Fig. 3.** Periodograms of the photometric and spectroscopic data of TOI-6628, TOI-3837, TOI-5027 and TOI-2328, respectively. The horizontal red dashed line corresponds to the GLS 1% False Alarm Probability and the blue vertical dashed line is marking the planet period.

### 2.2. Ground-based photometry

We used ground-based photometric time series obtained with sub-meter class telescopes to confirm the planetary nature of transiting candidates identified by TESS. These light curves are used to ensure that the transit-like feature occurs on the particular star of interest, and not on another star inside the photometric aperture of the TESS data. Given the low number of transits present in TESS data for the four systems presented in this study, these additional light curves are also crucial for refining the photometric ephemeris. Table 2.1 show the details of the ground-based observations follow-up observations of the planetary systems under study.

#### 2.2.1. Observatoire Moana

Observatoire Moana is a global network of telescopes with observing stations in Chile, Europe, and Australia. The observations cited here were made by OMES–CDK600 and OMES–RiDK500, both of which are located at El Sauce Observatory, Chile. OMES–CDK600 is an equatorial fork mounted 0.6 meter diameter Corrected Dall-Kirkham tele- scope and OMES–RiDK500 is a similarly mounted 0.5 meter Riccardi Dall-Kirkham instrument. Both telescopes are equipped with a range of filters, including the Sloan r filter, and Andor Ikon-L 936 cameras with deep depleted back illuminated 2K×2K CCD sensors. OMES–CDK600 has a pixel scale of 0.67" and a field of view of 0.4×0.4 degrees and OMES–RiDK500 has a pixel scale of 0.78" and a field of view of 0.6×0.6 degrees.

Reduction was performed using a dedicated automated pipeline which considers all sources available in the field and produces light curves and artificial comparison stars for all qualifying sources, derived from parameterised limits on variability and magnitude differentials, and an optimisation algorithm which also draws on the spatial separation between targets and other sources.

We observed two full transits of TOI-6628 *b* on the nights of 2021 June 19th using OMES-CDK600 and one transit during the night of 2024 March 15th using OMES-RiDK500.

We observed an almost-full transit of TOI-5027 *b* using OMES-RiDK500 on the night of 2024 March 21st.

We have observed half a transit of TOI-2328 *b*, including the planetary egress, on the night of 2021 October 23rd using OMES-CDK600.

#### 2.2.2. CHAT

The Chilean Hungarian Automated Telescope (CHAT) was a robotic facility installed at Las Campanas Observatory in Chile. CHAT consistss in a FORNAX 200 equatorial mount, and a 0.7m telescope coupled to a FLI ML-23042 CCD of 2048×2048 pixels, which delivers a pixel scale of 0.6"/pix. CHAT contains a set of $i'$, $r'$, and $g'$ passband filters.

We observed the transit of TOI-2328 *b* with CHAT on the night of 2019 December 26th with the $r'$ filter adopting an exposure time of 20s. We obtained 112 images of TOI-2328. CHAT data were processed with a dedicated pipeline that performs differential aperture photometry, where the optimal comparison sources and the radius of the photometric aperture are automatically selected (e.g., Espinoza et al. (2019); Jones et al. (2019); Jordán et al. (2019)). The light curve obtained is presented in the top second panel from left to right of Figure 7, and shows an egress for TOI-2328 *b* which confirms that the transit identified in the TESS data occurs in a region of 8" centered on TOI-2328.

#### 2.2.3. LCOGT

Las Cumbres Observatory global telescope network (LCOGT, Brown et al. (2013)) is a world-wide network of 1m telescopes, equipped with 4096×4096 SINISTRO cameras. The cameras have a pixel scale of 0.389 arcsec, resulting in a 26"×26" field of view and equipped with SDSS/PanSTARRS $u'g'r'i'zsYw$ filters. In addition, the LCOGT hosts a 0.35m telescope equipped with a QHY600 detector and a $i'$ filter, providing a pixel scale of 0.75 arcsec.

We observed a full transit of TOI-6628 *b* on 2024 May 26th with a cadence of 5 minutes using the LCOGT 0.35m telescope. We observed an almost full transit of TOI-5027 *b* on 2022 April 3rd with a cadence of 2 minutes, and one full transit and an egress of TOI-2328 *b* on 2020 September 24th and on 2021 August 15th, respectively, using the LCOGT 1m Telescopes.

All the transits were obtained using the $i'$ filter, except for the egress of TOI-2328 *b*, that was also observed using the $g'$ filter.

| Parameter | TOI-6628 *b* | TOI-3837 *b* | TOI-5027 *b* | TOI-2328 *b* |
|---|---|---|---|---|
| P (days) | $18.18424\pm^{0.00001}_{0.00001}$ | $11.88865\pm^{0.00003}_{0.00003}$ | $10.24368\pm^{0.00001}_{0.00001}$ | $17.10197\pm^{0.00000}_{0.00001}$ |
| t0 (days) | $2458602.7209\pm^{0.0009}_{0.0009}$ | $2459530.1610\pm^{0.0007}_{0.0008}$ | $2458649.4580\pm^{0.0010}_{0.0009}$ | $2458330.4894\pm^{0.0004}_{0.0003}$ |
| b | $0.814\pm^{0.016}_{0.017}$ | $0.728\pm^{0.032}_{0.051}$ | $0.926\pm^{0.026}_{0.017}$ | $0.688\pm^{0.022}_{0.018}$ |
| p | $0.100\pm^{0.001}_{0.001}$ | $0.082\pm^{0.002}_{0.002}$ | $0.109\pm^{0.013}_{0.006}$ | $0.102\pm^{0.000}_{0.000}$ |
| $e$ | $0.667\pm^{0.016}_{0.016}$ | $0.198\pm^{0.046}_{0.058}$ | $0.395\pm^{0.032}_{0.029}$ | $0.057\pm^{0.046}_{0.029}$ |
| $\omega$ (deg) | $215.4\pm^{2.4}_{2.3}$ | $285.3\pm^{12.4}_{11.7}$ | $288.6\pm^{3.4}_{3.4}$ | $148.7\pm^{14.1}_{9.3}$ |
| $\rho_*$ (kg/m$^3$) | $1299.3\pm^{89.3}_{95.0}$ | $723.6\pm^{37.3}_{38.4}$ | $1833.3\pm^{38.8}_{37.5}$ | $1806.1\pm^{42.1}_{35.4}$ |
| K (m/s) | $78.90\pm^{3.10}_{3.12}$ | $58.39\pm^{3.97}_{3.98}$ | $201.77\pm^{4.00}_{4.23}$ | $13.29\pm^{1.57}_{1.53}$ |
| $i$ (deg) | 88.17±0.13 | 87.98±0.17 | 88.18±0.10 | 88.66±0.08 |
| $m_p$ (m$_J$) | $0.75\pm^{0.06}_{0.06}$ | $0.59\pm^{0.06}_{0.06}$ | $2.01\pm^{0.13}_{0.13}$ | $0.16\pm^{0.02}_{0.02}$ |
| $a$ (AU) | $0.133\pm^{0.007}_{0.007}$ | $0.098\pm^{0.005}_{0.005}$ | $0.093\pm^{0.004}_{0.004}$ | $0.127\pm^{0.005}_{0.006}$ |
| $R_p$ ($R_J$) | $0.98\pm^{0.05}_{0.05}$ | $0.96\pm^{0.05}_{0.05}$ | $0.99\pm^{0.07}_{0.12}$ | $0.89\pm^{0.04}_{0.04}$ |
| $\rho_p$ (g/cm$^3$) | $0.97\pm^{0.16}_{0.16}$ | $0.85\pm^{0.16}_{0.16}$ | $2.77\pm^{0.90}_{0.90}$ | $0.29\pm^{0.05}_{0.05}$ |
| $T_{eq}$ (K) | $836\pm^{22}_{23}$ | $1182\pm^{30}_{31}$ | $1056\pm^{23}_{24}$ | $842\pm^{20}_{21}$ |

**Table 3.** Orbital parameters.

All LCOGT science images were calibrated by the standard LCOGT BANZAI pipeline (McCully et al. 2018), and photometric measurements were extracted using AstroImageJ (Collins et al. 2017).

### 2.3. Hazelwood Observatory

The *Hazelwood* Observatory is a backyard observatory located in Victoria, Australia that host a 0.32m Planewave CDK F/8 telescope equipped with a SBIG STT3200 2.2k×1.5k CCD, providing a 20′×13′ FoV and plate scale of 0.55" per pixel. The system is equipped with a filter-wheel with B, V, Rc, Ic, $g'$ , $r'$, $i'$ , and $z'$ filters.

We observed an ingress of TOI-6628 *b* the night of 2024 April 20th and an egress of TOI-2328 *b* the night of 2021 February 8th, both using the Rc filter.

### 2.4. KeplerCam

KeplerCam is a 4K×4K CCD camera installed in the 1.2m Telescope at the Fred Lawrence Whipple Observatory (FLWO) in Mount Hopkins, Arizona, USA. The system provides a pixel scale of 0.672" per pixel at 2×2 pixel binning.

We obtained an egress of TOI-3837 *b* the night of 2022 April 27th with a cadence of 30 seconds using the Sloan $i'$ filter.

### 2.5. Spectroscopy

We obtained precision radial velocities (RVs) for our four targets from three different instruments. The main goal is to search for RV variations consistsent with the planetary hypotheses of the transiting planets. All FEROS and HARPS data obtained for this study were processed with the `ceres` pipeline (Brahm et al. 2017). `ceres` performs all the reduction steps to generate a continuum-normalized, wavelength-calibrated, and optimally extracted spectrum from raw data. `ceres` also computes precision radial velocities and bisector spans through the cross-correlation technique. The bisector analysis of the cross-correlation function (CCF) profile can be used as an indicator of phospheric activity (e.g. (Queloz et al. 2001)). We employ G2-type binary mask as a template for computing the CCF of the spectra of our four candidates.

#### 2.5.1. FEROS

We use the Fiber-fed Extended Range Optical Spectrograph (FEROS; Kaufer et al. (1999)) installed in the ESO-MPG 2.2m telescope at La Silla Observatory, Chile, for spectroscopic follow-up. With an average resolving power of $R$ ~48,000, it can measure stellar RVs with an accuracy of the order of few m/s, which is suitable for exoplanet searches.

Exposure times ranged from 900 $s$ to 1500 $s$. The instrument is equipped with two fibers, allowing us to simultaneously observe the spectra of a secondary source, for two purposes: to subtract the sky background or for simultaneous wavelength calibration. In our case, the latter was the observing mode chosen for our observations.

We obtained 3 RVs for TOI-6628, between the 23th of Februrary and the 22th of July of 2021, with a mean uncertainty of 9.3 m/s in the RVs. The observations were done in the object-calibration mode, with an exposure time of 25 minutes, resulting in an average $S/N$ = 53. Table A.1 show the RVs obtained for TOI-6628.

We obtained 8 RVs for TOI-3837, between the 5th of January of 2021 and the 8th of March of 2021. The mean uncertainty of the RVs is 10.1 m/s, obtained in object calibration mode with an exposure time of 15 minutes. The average (S/N) of the observations is 70. Table A.2 show the RVs obtained for TOI-3837.

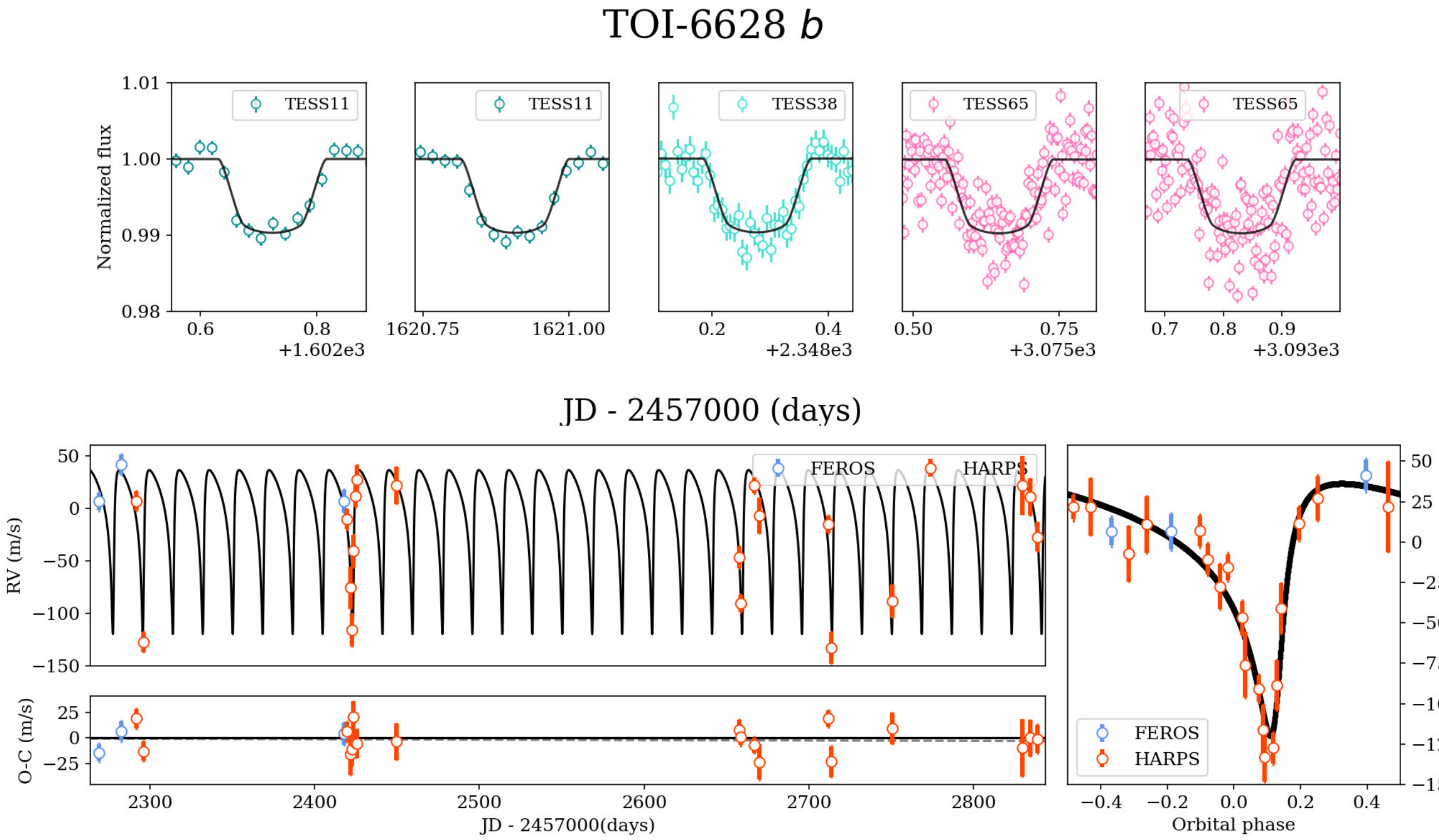

**Fig. 4.** Lightcurves and RV data of TOI-6628. We have shown the best `juliet` joint fit model in black. FEROS RVs are shown in blue and HARPS RVs are shown in red.

| Model | TOI-6628 | TOI-3837 | TOI-5027 | TOI-2328 |
|---|---|---|---|---|
| a) | 6573.95 | 3017.2 | 1378.7 | 67670.3 |
| b) | 8658.2 | 5695.7 | 4116.1 | 73272.9 |
| c) | 8733.7 | 5699.7 | 4151.2 | 73294.1 |
| d) | 8718.6 | 5682.7 | 4134.6 | 73251.1 |

**Table 4.** Log-evidence values of the four different models we tested for the planets here presented. a) No planet, b) 1 planet, $e = 0$, c) 1 pl., $e \neq 0$, d) 1 pl., $e \neq 0$, with linear trend.

We obtained 23 RVs for TOI-5027, between the 13th of April of 2022 and the 5th of June of 2023. The mean uncertainty of the RVs is 11.2 m/s, obtained in object-calibration mode with an exposure time of 20 minutes. The average (S/N) of the observations is 80. Table A.3 show the RVs obtained for TOI-5027.

We obtained 22 RVs for TOI-2328, between the 11th of July of 2022 and the 23rd of November of 2022, with a mean RV uncertainty of 9.2 m/s. The observations were obtained in object-calibration mode with an exposure time of 20 min, reaching an average (S/N) of 50. Table A.4 show the RVs obtained for TOI-2328.

### 2.5.2. HARPS

The High Accuracy Radial velocity Planet Searcher (HARPS; Mayor et al. (2003)) is a high-resolution and stabilized spectrograph fiber-fed by the Cassegrain focus of the ESO 3.6 m telescope of the ESO La Silla Observatory, in Chile. It covers the spectral region of 380–690 nm, with a resolving power of R = 115,000, making it suitable for RV follow-up of exoplanets with low semi-amplitude of the RV variations.

We obtained 20 RVs for TOI-6628, between the 18th of March of 2021 and the 11th of September of 2022. The mean uncertainty of the RVs is 13.5 m/s, obtained with an exposure time of 30 min and an average (S/N) = 25. Table A.1 show the RVs obtained for TOI-6628.

We obtained 7 RVs for TOI-3837, between the 4th of February of 2021 and the 21st of March of 2021. The mean uncertainty of the RVs is 9.2 m/s, obtained in object calibration mode with an exposure time of 20 minutes. The average (S/N) of the observations is 20. Table A.2 show the RVs obtained for TOI-3837.

We obtained 22 RVs for TOI-2328, between the 14th December 2019 and the 26th of July of 2021. The mean uncertainty of the RVs is 8.3 m/s, obtained in object calibration mode with an exposure time of 20 min, reaching an average (S/N) = 25. Table A.4 show the RVs obtained for TOI-2328.

### 2.5.3. PFS

We monitored TOI-2328 with the Planet Finder Spectrograph (PFS; Crane et al. (2006, 2008, 2010)) installed at the 6.5 m Magellan/Clay telescope at Las Campanas Observatory. The target was observed using the iodine gas absorption cell of the instrument during four different observing runs between 2020 November 3 and 2021 November 16, adopting an exposure time of 1200 s and using a 3×3 CCD binning mode to minimize readout noise. We also observed TOI-2328 without the iodine cell in order to generate a template for computing the RVs, which were derived following the methodology of Butler et al. (1996). We obtained 12 RVs from PFS with a mean uncertainty of 5.7 m/s. The PFS RVs are presented in Table A.4.

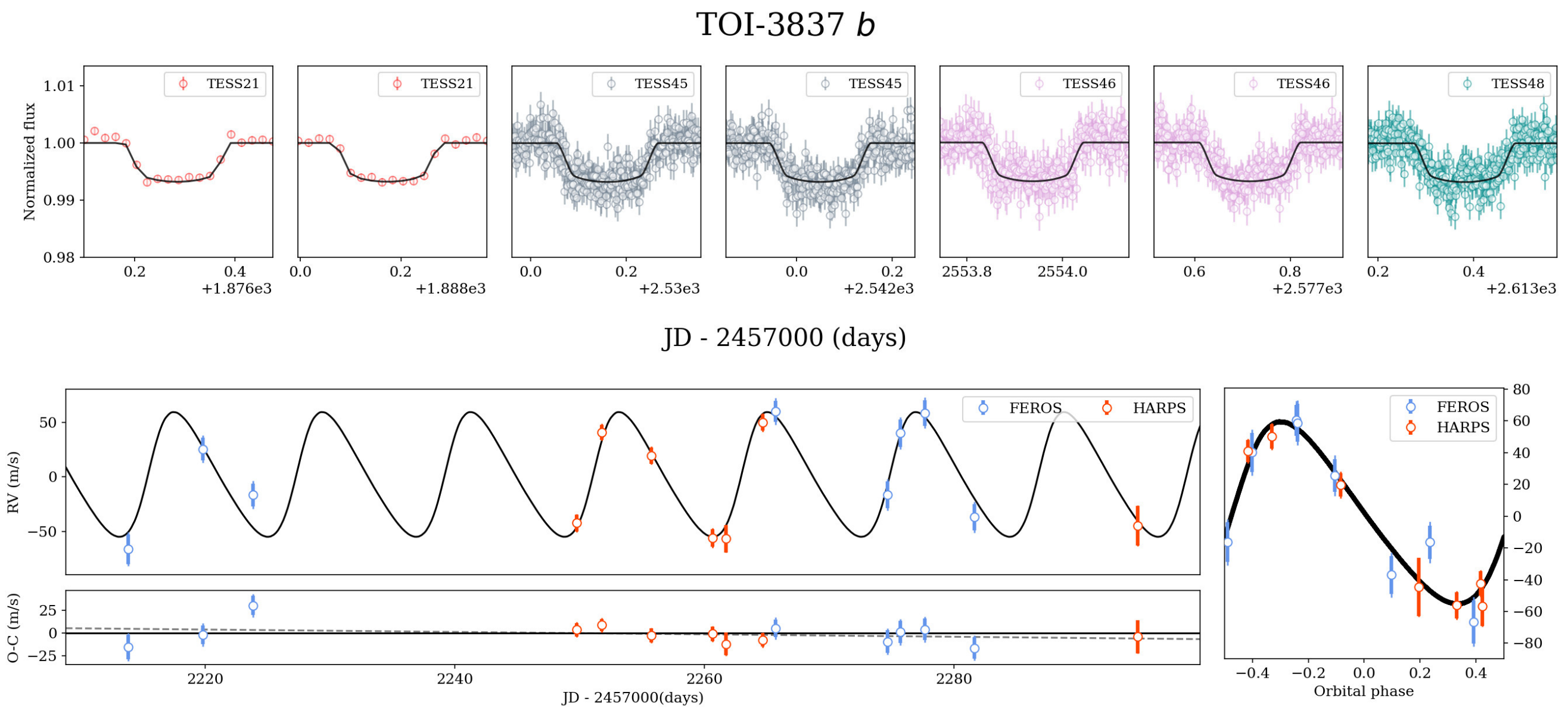

**Fig. 5.** Lightcurves and RV data of TOI 3837. We have shown the best `juliet` joint fit model in black. FEROS RVs are shown in blue and HARPS RVs are shown in red.

### 2.6. High resolution imaging

We use the optical speckle imaging technique to obtain high resolution images for two of our targets, TOI-3837 and TOI-5027. Both targets were observed with the HRCam of the Southern Astrophysical Research (SOAR) telescope (Tokovinin 2018; Ziegler et al. 2020), and TOI-3837 was also observed by the Sternberg Astronomical Institute (SAI) 2.5m telescope (Shatsky et al. 2020). The SOAR observations for TOI-3837 and TOI-5027 were obtained on the 6th of June of 2022 and on the 22th of April of 2022, respectively. The SAI-2.5m telescope observation for TOI-3837 was obtained on the 28th of November of 2022.

No nearby companion was found nearby TOI-5027 at a contrast of Δmag=6 at 1”. However, we identified a faint companion ($\Delta I$=4) nearby TOI-3837 at 0.4” (see Fig.2). The flux associated with the neighboring star might impact the determination of the physical parameters of the host star and the planetary transit parameters by diluting the transits. These effects are taken into account in the analysis of the TOI-3837 *b* system presented in section 4.

## 3. Stellar parameters

Table 1 show the main stellar parameters of our targets. The sky positions right ascension (RA) and declination (Dec) are obtained from the Gaia DR2 (Gaia Collaboration et al. 2018), as well as the proper motions and the stellar parallax. We have shown also the TESS magnitude $T$ (Ricker et al. 2015), the V and B magnitudes (Munari et al. 2014), the G magnitude (Gaia Collaboration et al. 2018), and the J, H and K magnitudes (Skrutskie et al. 2006). Using these data, we derived the stellar properties of the four host stars using the Zonal Atmospheric Stellar Parameters Estimator (`zaspe`, Brahm et al. (2017)) code, following the iterative procedure described by Brahm et al. (2019b).

First, we create a high SNR stellar template from the co-added stellar spectra observed for each star. This stellar template is then compared to a grid of synthetic spectra from the ATLAS9 models (Castelli & Kurucz 2003). After that, `zaspe` derives an initial set of stellar atmospheric parameters, namely, effective temperature ($T_{eff}$), surface gravity (log$g$), metalicity ([Fe/H]) and projected rotational velocity ($v$ sin$i$). Then, the star's physical parameters, mass ($M_*$), radius ($R_*$), luminosity ($L_*$), density $\rho_*$, age and visual extinction along the line of sight ($A_V$), are obtained by comparing the stellar broad-band photometry with the synthetic magnitudes from the `parsec` stellar evolutionary models (Bressan et al. 2012). Absolute magnitudes are obtained using Gaia (Gaia Collaboration et al. 2018) parallaxes.

We use the values of [Fe/H] and $T_{eff}$ as priors to the calculation of the physical parameters with `parsec`. While the value of [Fe/H] remains fixed, $T_{eff}$ is used an initial value in the calculation. The procedure is repeated in an iterative way until it converges.

Following Tayar et al. (2022), we add systematic uncertainties of 2% to the stellar luminosity, an uncertainty of 4% to the stellar radius, an uncertainty of 5% to the stellar mass, and an uncertainty of 20% to the stellar age. The resulting stellar parameters for the four stars here presented are listed in Table 1.

### 3.1. Activity indices

We computed stellar activity indices from the reduced data using `ceres`. Besides the bisectors of the CCF, we computed the Hα, NaII, HeI and CaII indices, which can be used as tracers of chromospheric activity (Andretta et al. 2005). For the Hα index we follow the the definition from Boisse et al. (2009). For the NaII and HeI indices we follow the defition from Gomes da Silva et al. (2011). For the CaII index we follow the defition from Duncan et al. (1991). The activity indices for TOI-6628, TOI-3837, TOI-5027 and TOI-2328 are listed in tables A.1, A.2, A.3 and A.4, and shown against the RVs in figures B.1, B.2, B.3 and B.4, respectively.

## 4. Analysis and results

We analyzed the transits and RV time series of the four systems to search for periodical signals. We computed the Box Least-Square periodogram (BLS; Kovács et al. (2002)) in the photometry time series by combining the data of the different instru-ments and epochs for each target. To search for periodicities in the RVs, and in the corresponding activity indices, we computed the generalized Lomb–Scargle periodograms (GLS; Zechmeister & Kürster (2018)) by combining the radial velocities of different instruments for each system.

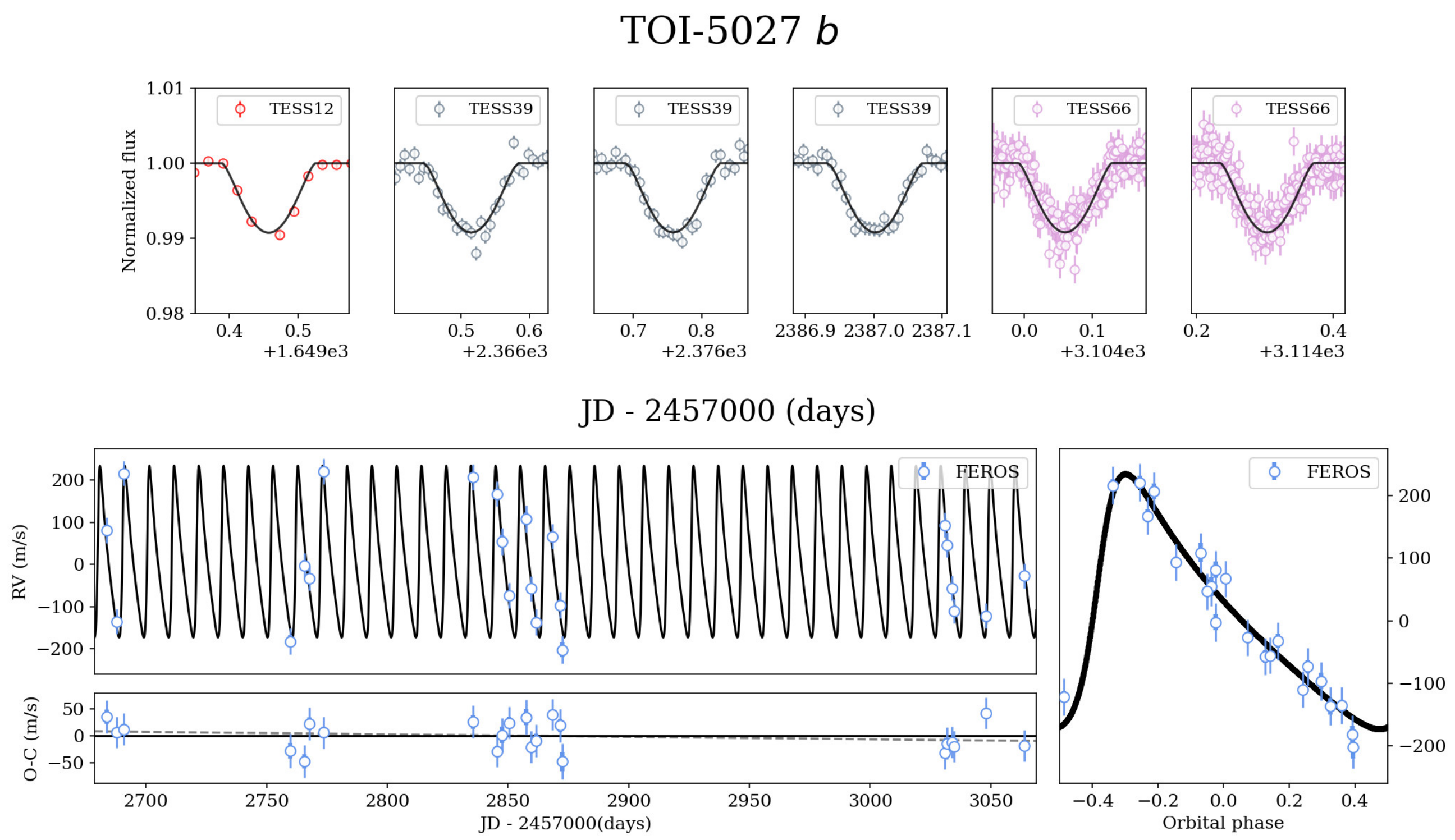

**Fig. 6.** Lightcurves and RV data of TOI-5027. We have shown the best `juliet` joint fit model in black. FEROS RVs are shown in blue.

Figure 3 show the BLS and GLS for the four targets under study, for the photometric and spectroscopic data, respectively.

With the only exception of TOI-6628, all the targets show a significant peak in the RVs GLS that coincides with the peak observed in the photometry BLS. In the case of TOI-6628, we observe a peak at the period of the RVs GLS peak, but is not significant over the 1% false alarm probability (FAP). One possible explanation for this is the large eccentricity of the predicted planetary orbit, which could affect the accuracy of the GLS to infer periods from such eccentric RV dataset.

We do not observe any significant peaks in the RV residuals, neither in the activity indices.

The BLS and GLS analysis reveal periodicities at 18.18 d, 11.89 d, 10.24 d and 17.1 d, for TOI-6628, TOI-3837, TOI-5027 and TOI-2328, respectively. These values are used as priors for the period $P$ in the planetary orbit modeling.

### 4.1. Orbit modeling

We performed a joint analysis of the photometric and spectroscopic data using `juliet` (Espinoza et al. 2019), a software package that uses Bayesian inference and nested sampling algorithms to find the most likely model to the data. It is based on `radvel` (Fulton et al. 2018) for the RV modeling and on `batman` (Kreidberg 2015) for the transit modeling. We employ the `dynesty` (Speagle 2020) package, included in `juliet`, to sample the posterior distributions.

`juliet` models define two types of parameters: orbital and instrumental. The orbital parameters are: the orbital period $P$, the time of the transit $t_0$, the planet to star radius $p$ ($p = R_p/R_*$), the impact parameter of the orbit $b$, the stellar density $\rho_*$, the eccentricity $e$, the argument of periastron passage $\omega$ of the orbit and the planet RV semi-amplitude $K$.

The instrumental parameters describe the properties of each dataset. For each photometric instrument the parameters are: the dilution factor $m_D$, the relative flux offset $m_{Flux}$, the photometric jitter value $\sigma_w$ and the limb-darkening parameters $q_1$ and $q_2$. For each spectroscopic instrument the parameters are: the systemic RV of a given instrument $\mu$ and the RV jitter $\sigma_w$. In some cases we incorporate a linear term $\theta_0$ to account for the effect of airmass in ground-based lightcurves.

From the orbital parameters one can derive the orbital inclination $i$, the planet mass $M_P$, the planet radius $R_P$, the semi-major axis $a$, the planet density $\rho_P$ and the planet surface equilibrium temperature $T_{eq}$ (assuming zero albedo and uniform planet surface temperature; Méndez & Rivera-Valentín (2017)). We built posterior distributions for the derived parameters using the orbital parameter posteriors, from which we estimate their uncertainties.

We tested four different models: i) no planet, ii) a single planet with zero eccentricity, iii) a single planet with non-zero eccentricity and iv) a single planet with non-zero eccentricity and a linear trend. By comparing the Bayesian log-evidence we can quantify the validity of the different models given the data. For model i) we only fit the instrumental parameters, assuming that the observed signals in the photometry and in the RVs is produced by instrumental noise. For model ii), $e$ is fixed and set to 0 and $\omega$ is fixed and set to 90 deg. For model iii) all the orbital and instrumental parameters are fitted, and for model iv) we add a linear trend to the RV model, which is parametrized by its slope and intercept.

Table 4 show the log-evidence values for the different models, showing that the model of the single planet orbiting its host star in an eccentric orbit is the most likely model in all four cases. However, we note that in the case of TOI-3837, the Δlog-evidence<5. While for the best-fit model $e = 0.198$, more RV should provide better constraints to the orbit, as the Δlog-evidence suggest that a circular orbit could also provide a good description of the data. We observe a trend when fitting a line to the RV residuals of TOI-3837 *b*, TOI-6628 *b* and TOI-5027 *b*, however adding a linear trend to the RV model does not improve the results.

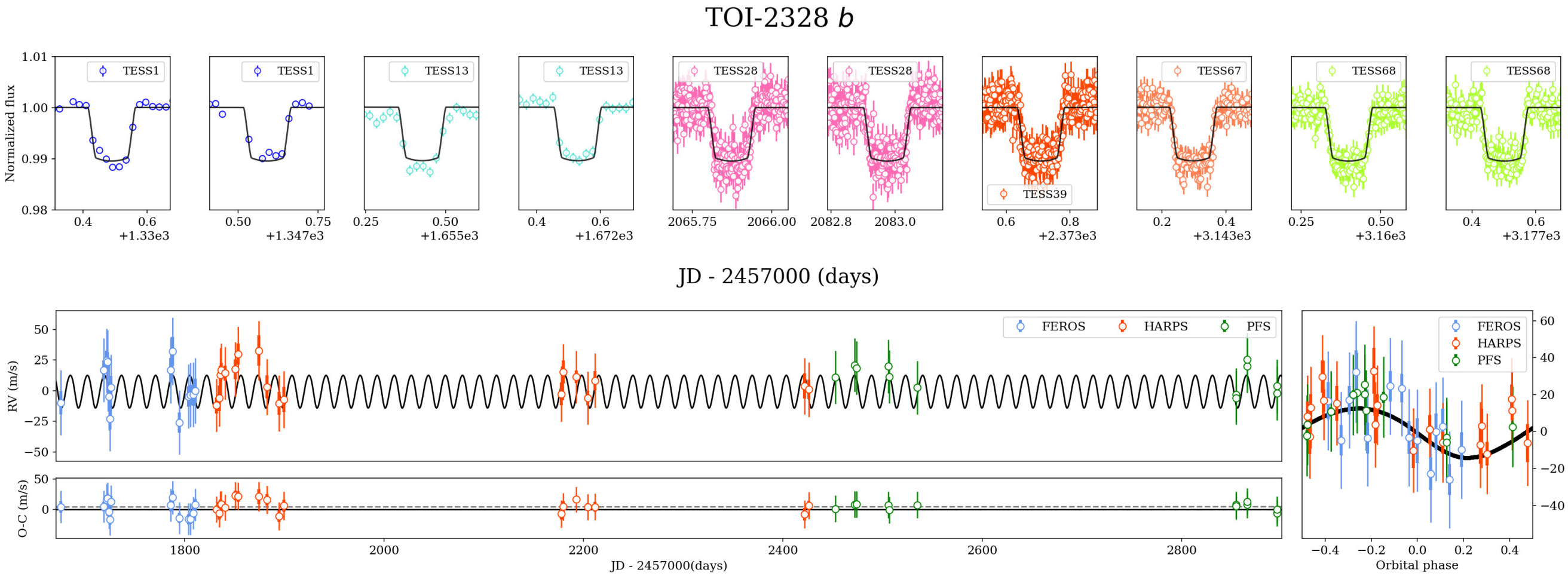

**Fig. 7.** Lightcurves and RV data of TOI-2328. We have shown the best `juliet` joint fit model in black. FEROS RVs are shown in blue, HARPS RVs are shown in red and PFS RVs are shown in green.

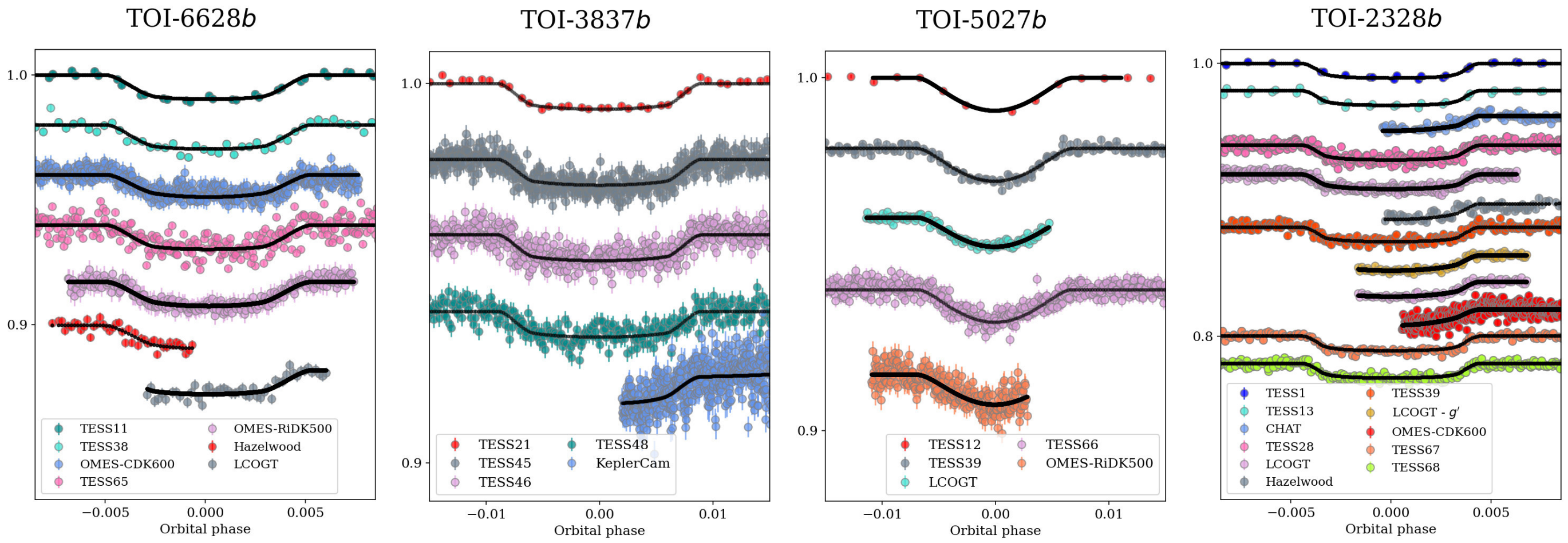

**Fig. 8.** Phase folded lightcurve of TOI-6628, TOI-3837, TOI-5027 and TOI-2328. The best model is shown in black.

The tables 6, 7, 8 and 9 show the priors adopted for the parameters of model iii), along with their best-fit values, for TOI-6628, TOI-3837, TOI-5027 and TOI-2328, respectively.

#### 4.1.1. TOI-6628 *b*

The dataset used for the global modelling of TOI-6628 consists of 3 TESS lightcurves, 1 lightcurve from OMES-CDK600, 1 lightcurve from OMES-RiDK500, 1 lightcurve from *Hazelwood* Observatory, 1 lightcurve from LCOGT, 3 FEROS RVs and 20 HARPS RVs. The lightcurves from TESS sectors 11 and 38 are from the PDC-SAP data, while the one from sector 65 are from the QLP data. For the transit modelling we have adopted a quadratic limb-darkening law. FEROS and HARPS RVs are modeled with a keplerian eccentric orbit. Instrumental parameters consider distinct flux and RV offsets and jitters as free parameters. We have included a linear term to model the effect of airmass in the *Hazelwood* lightcurve. Table 6 show the prior distributions, along with the best-fit parameters for the adopted model of TOI-6628 *b*. The best-fit model is shown in black in Fig. 4, and the phase-folded lightcurves are shown in Fig. 8.

According to our analysis TOI-6628 is a WJ with a mass $M_P = 0.75\pm^{0.06}_{0.06}$ $M_J$ and a radius $R_P = 0.98\pm^{0.05}_{0.05}$ $R_J$, resulting a mean density $\rho_p = 0.97\pm^{0.16}_{0.16}$ (g/cm$^3$). It orbits its host star every 18.18424±0.00001 days in an eccentric orbit with $e = 0.667\pm^{0.016}_{0.016}$ and has an time-averaged equilibrium temperature of $T_{eq} = 836$ K$\pm^{22}_{23}$.

#### 4.1.2. TOI-3837 *b*

The dataset used for the global modelling of TOI-3837 *b* consists of 4 TESS PDC-SAP lightcurves, 1 KeplerCam lightcurve, 8 FEROS RVs and 7 HARPS RVs. For the transit modelling we have adopted a quadratic limb-darkening law. FEROS and HARPS RVs are modeled with a keplerian eccentric orbit. Instrumental parameters consider distinct flux and RV offsets and jitters as free parameters. Table 7 show the prior distributions, along with the best-fit parameters for the adopted model of TOI-3837. The best-fit model is shown in black in Fig. 5, and the phase-folded lightcurves are shown in Fig. 8.

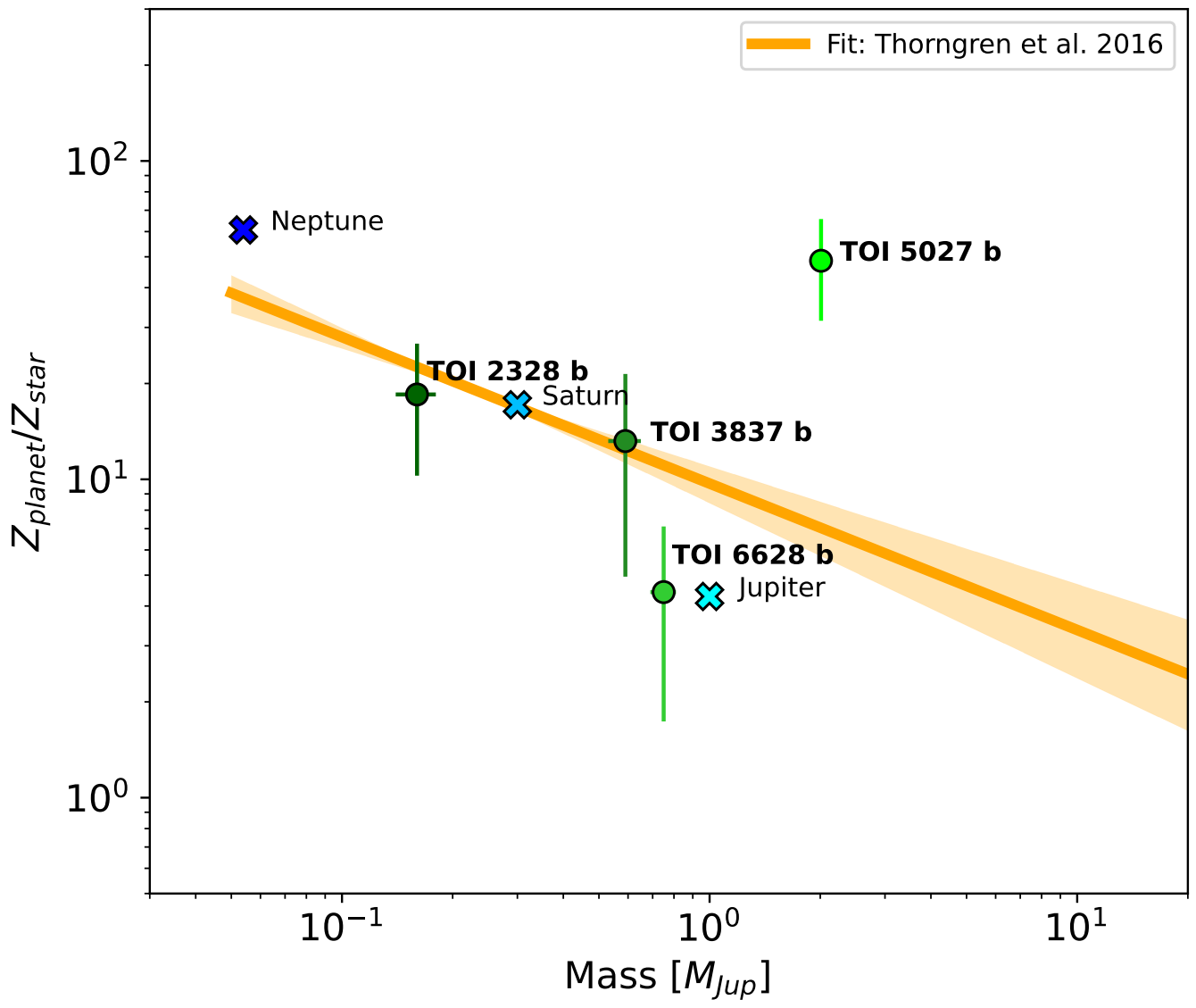

**Fig. 9.** Mass-bulk metallicity diagram. To compare the composition relative to the host star of the four newly discovered warm gas giants, we show the Solar System gas and ice giants, as well as the fit obtained by Thorngren et al. (2016) for a sample of extrasolar warm gas giants.

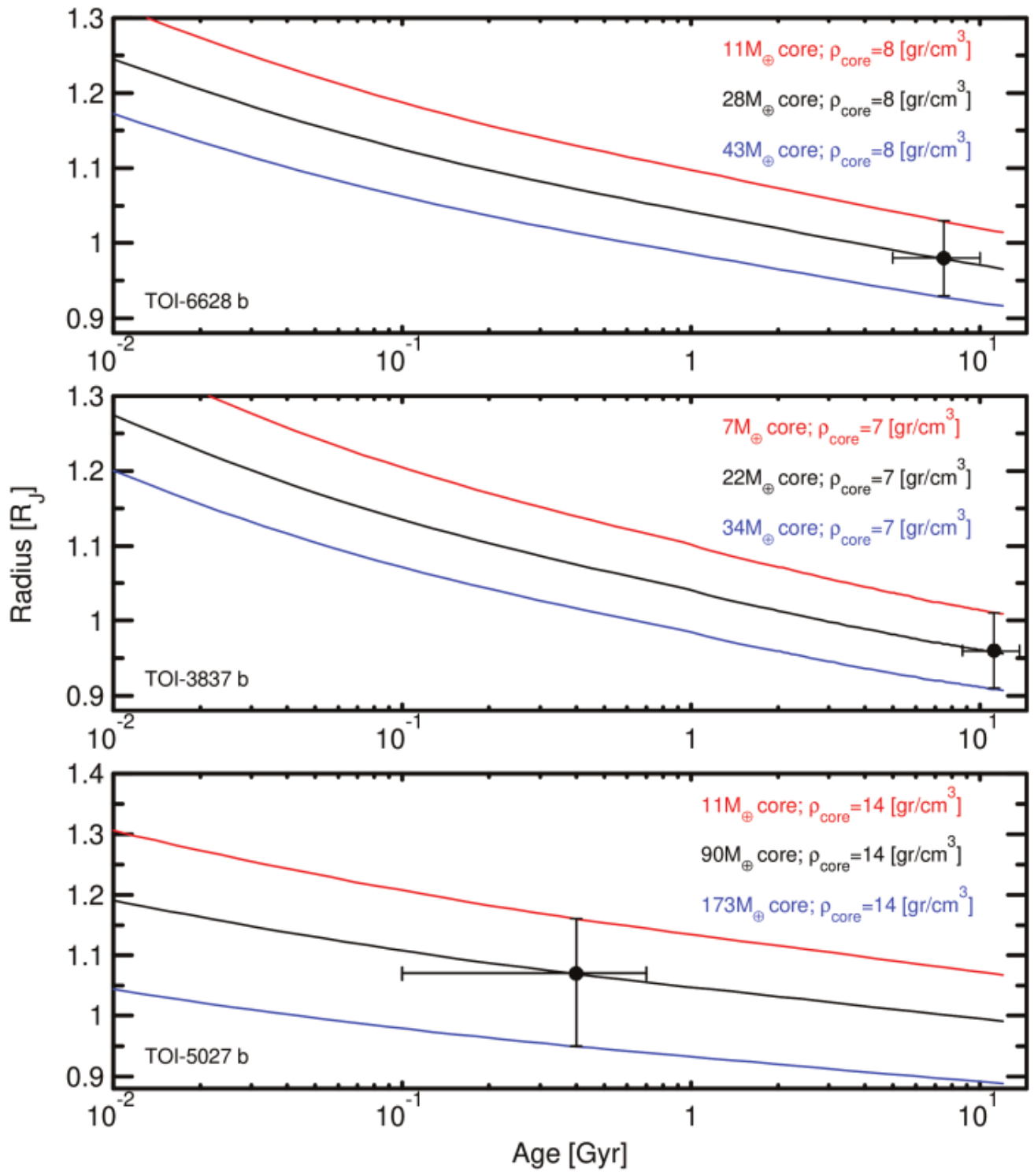

**Fig. 10.** Position of TOI-6628 *b*, TOI-3837 *b* and TOI-5027 *b* in the age-radius diagram (black dot in the upper, middle and lower panel, respectively). Planet evolutionary models with different core mass and density are over-plotted.

We identified a close-by companion to TOI-3837 at 0.4” in the high resolution images provided by SOAR and SAI (see Fig 2). Therefore, we attempted to model TOI-3837 as a blended stellar eclipsing binary system following the methods of Hartman et al. (2019). Here we assume that the fainter of the two resolved components is an eclipsing binary, and we consider both the case where it is physically associated to the brighter resolved star (i.e., has the same age, distance and metallicity), as well as the case where the brighter resolved star is not associated with the fainter object. In both cases we assume that all of the stars have physical properties that are constrained by the MIST 1.2 stellar evolution models, and we model the photometric light curves, calibrated broad-band photometric measurements, the astrometric parallax measurement, and the spectroscopic temperature and metallicity (which we assume are measured for the brighter of the two resolved objects). We assume that the two components resolved in the high-spatial resolution imaging are fully blended in all of the light curves and catalog broad-band photometry measurements, while the magnitude differences between the two components that are measured by the high-spatial resolution imaging are treated as additional observables to be fitted by the model. We allow the eccentricity and argument of periastron to vary in the fit (using $\sqrt{e}\cos\omega$ and $\sqrt{e}\sin\omega$ as the adjusted parameters), but do not include the RVs in this analysis.

For comparison, we also perform a similar modelling of all of these data treating the brighter of the resolved components as a star with a transiting planet, while assuming the fainter component is an individual star that is physically bound to the brighter component. Here we first perform the analysis including the RV observations also in the model, and then perform a second analysis excluding the RV observations, and fixing the eccentricity and argument of periastron to the values determined when the RVs were included in the fit.

We find that modelling the observations as a transiting planet system in a resolved stellar binary provides a significantly better (lower $\chi^2$) fit to the data than the blended stellar eclipsing binary models. The best-fit transiting planet model, excluding the RVs, has a $\chi^2$ that is lower by 188.2 than the best-fit blended stellar eclipsing binary model, despite being a less complicated model with fewer varied parameters. Moreover, the significant RV variation, consistsent with a planetary mass companion, and the lack of a corresponding BIS variation also are strong indicators that the object is a transiting planet system, and not a blended stellar eclipsing binary.

Therefore, our analysis indicates that TOI-3837 is a WJ with a mass $M_P = 0.59\pm^{0.06}_{0.06}$ $M_J$ and a radius $R_P = 0.96\pm^{0.05}_{0.05}$ $R_J$, yielding a mean density $\rho_p = 0.85\pm^{0.16}_{0.16}$ (g/cm$^3$). It orbits its host star every 11.88865±0.00003 days in a low eccentricity orbit with $e = 0.198\pm^{0.046}_{0.058}$ and has an time-averaged equilibrium temperature of $T_{eq} = 1182\pm^{30}_{31}$ K.

### 4.1.3. TOI-5027 *b*

The dataset used for the global modelling of TOI-5027 *b* consists of three TESS lightcurves, one LCOGT lightcurve, one OMES-RiDK500 lightcurve and 16 FEROS RVs. For the transit modelling we have adopted a quadratic limb-darkening law. FEROS RVs are modeled with a keplerian eccentric orbit. Table 8 show the prior distributions, along with the best-fit parameters for the adopted model of TOI-5027 *b*. The best-fit model is shown in black in Fig. 6, and the phase-folded lightcurves are shown in the center-right panel in Fig. 8.

Our analysis indicates that TOI-5027 *b* is a WJ with a mass $M_P = 2.01\pm^{0.13}_{0.13}$ $M_J$ and a radius $R_P = 0.99\pm^{0.07}_{0.12}$ $R_J$, yielding a mean density $\rho_p = 2.77\pm^{0.90}_{0.90}$ (g/cm$^3$). It orbits its host star every 10.24368±0.00001 days in an eccentric orbit with $e$ = 0.395$\pm^{0.032}_{0.029}$ and has an time-averaged equilibrium temperature of $T_{eq}$ = 1056 K$\pm^{23}_{24}$.

| | CMF | | log(Fe/H)$_p$ | | $Z_{env}$ | | $Z_p$ | | $Z_p/Z_*$ | |
|---|---|---|---|---|---|---|---|---|---|---|
| Planet | GASTLI | MESA | GASTLI | MESA | GASTLI | MESA | GASTLI | MESA | GASTLI | MESA |
| TOI-6628 *b* | $0.11^{+0.11}_{-0.08}$ | $0.12^{+0.06}_{-0.07}$ | $-0.55^{+1.04}_{-1.00}$ | – | $0.01^{+0.05}_{-0.01}$ | – | $0.14^{+0.11}_{-0.08}$ | $0.12^{+0.06}_{-0.07}$ | 4.4±2.7 | $3.5^{+1.7}_{-3.0}$ |
| TOI-3837 *b* | $0.08^{+0.08}_{-0.05}$ | $0.12^{+0.06}_{-0.08}$ | $-0.77^{+0.95}_{-0.84}$ | – | $0.01^{+0.02}_{-0.01}$ | – | $0.10^{+0.08}_{-0.06}$ | $0.12^{+0.06}_{-0.08}$ | 13.2±8.3 | $15.0^{+7.5}_{-10.0}$ |
| TOI-5027 *b* | $0.47^{+0.19}_{-0.24}$ | $0.14^{+0.13}_{-0.13}$ | $-0.20^{+1.59}_{-1.29}$ | – | $0.01^{+0.28}_{-0.01}$ | – | $0.54^{+0.17}_{-0.21}$ | $0.14^{+0.13}_{-0.13}$ | 48.6±17.1 | $13.2^{+12.2}_{-12.2}$ |
| TOI 2328 *b* | $0.29^{+0.20}_{-0.16}$ | – | $-0.30^{+1.16}_{-1.18}$ | – | $0.01^{+0.09}_{-0.01}$ | – | 0.34±0.17 | – | 18.5± 8.2 | – |

**Table 5.** Interior modelling parameters of TOI-6628 *b*, TOI-3837 *b*, TOI-5027 *b* and TOI-2328 *b*, calculated using GASTLI and MESA. GASTLI mean and 1$\sigma$ uncertainties of the free parameters are derived from the posterior distribution functions obtained by the interior structure MCMC retrieval. These parameters correspond to the core mass fraction (CMF), the atmospheric metallicity (× solar), the envelope metal mass fraction $Z_{env}$, and the total bulk metal mass fraction $Z_{planet}$. MESA mean and 1$\sigma$ uncertainties of the interior model parameters are calculated following Jones et al. (2024).

#### 4.1.4. TOI-2328 *b*

The dataset used for the orbit modelling of TOI-2328 *b* consists of 6 TESS lightcurves, 1 CHAT lightcurve, 1 LCOGT lightcurve, 1 *Hazelwood* lightcurve, 1 OMES-CDK600 lightcurve, 16 FEROS RVs, 12 HARPS RVs and 12 PFS RVs. The TESS lightcurves from sectors 1 and 13 are from the 30 min cadence TESS-SPOC data, while the TESS lightcurves from sectors 27, 28, 39, 67 and 68 are from the 2 min cadence TESS-SPOC PDC-SAP data (Caldwell et al. 2020). For the transit modelling we have adopted a quadratic limb-darkening law. We have included a linear term to model the effect of airmass in the lightcurve from *Hazelwood*. FEROS, HARPS and PFS RVs are modeled with a keplerian eccentric orbit. Table 9 show the prior distributions, along with the best-fit parameters for the adopted model of TOI-2328 *b*. The best-fit model is shown in black in Fig. 7, and the phase-folded lightcurves are shown in the bottom right panel in Fig. 8.

TOI-2328 *b* is a Saturn-like planet with a mass $M_P$ = 0.16$\pm^{0.02}_{0.02}$ M$_J$ and a radius R$_P$ = 0.89$\pm^{0.04}_{0.04}$ R$_J$, with a mean density $\rho_p$ = 0.29$\pm^{0.05}_{0.05}$ (g/cm$^3$). It orbits its host star every 17.10197±0.00001 days in an orbit with $e$ = 0.057$\pm^{0.046}_{0.029}$ and has an time-averaged equilibrium temperature of $T_{eq}$ = 842$\pm^{20}_{21}$ K.

## 5. Interior modelling

We generate interior models with the GAS gianT modeL for Interiors (GASTLI, Acuña et al. 2024, 2021). The equations of hydrostatic equilibrium, adiabatic temperature, Gauss's theorem and conservation of mass are solved along a 1D grid that represents the planet radius from the center up to the interior boundary at a pressure of 1000 bar. The grid is stratified in two layers: a core composed of a 1:1 rock and water mixture, and an envelope constituted by H/He and water. The amount of water in the envelope is determined by the envelope metallicity, log([Fe/H]$_P$), which is a free parameter. We use up-to-date equations of state (EOS) to compute the density of silicates (Lyon 1992; Miguel et al. 2022), water (Mazevet et al. 2019; Haldemann et al. 2020) and H/He (Chabrier & Debras 2021; Howard & Guillot 2023). The interior model is coupled to a grid of cloud-free, self-consistsent atmospheric models obtained with petitCODE (Mollière et al. 2015, 2017), to calculate the atmospheric temperature. The conversion from atmospheric metallicity to metal mass fraction, $Z_{env}$, which is required to calculate the density of the H/He-water mixture, is done by assuming chemical equilibrium with easyCHEM (Mollière et al. 2017). We ensure convergence of the interior radius and boundary temperature with an iterative algorithm (Acuña et al. 2021).

Finally, the total radius is calculated by adding the atmospheric thickness at the transit pressure (Lopez & Fortney 2014; Grimm et al. 2018; Mousis et al. 2020, 20 mbar), to the converged interior radius. The age of the planet of each interior model is calculated by solving the equation of thermal cooling, $L = 4\pi\sigma R^2 T^4_{int}$, where $T_{int}$ is the internal (or intrinsic) temperature (Fortney et al. 2007; Acuña et al. 2024). This, and the planet's equilibrium temperature, are free parameters in our atmospheric grid. In addition to the planet mass and the envelope metallicity, the core mass fraction (CMF) is a free variable that parametrizes the composition of the planet. The bulk metal mass fraction of the planet, which comprises the metals contained both in the core and the envelope, can then be obtained as $Z_{planet}$ = CMF + (1-CMF) × $Z_{env}$.

We use a Markov chain Monte Carlo (MCMC) algorithm (`emcee`, Foreman-Mackey et al. (2013)) to sample the posterior distribution. We adopt uniform priors for CMF (0 to 0.99), log(Fe/H) (-2 to 2, equivalent to subsolar to 100 × solar) and $T_{int}$ (50 to 400 K), while the planet mass, $M_{\rm P}$, follows a Gaussian prior according to its observed mean and uncertainties (see Table 2.2.2). The log-likelihood is calculated with the squared residuals of the observable parameters (mass, radius and age) as indicated in Dorn et al. (2015); Acuña et al. (2021). Our grid of atmospheric models spans an equilibrium temperature from 100 to 1000 K, hence for TOI-5027 *b* and TOI-3837 *b*, we adopt $T_{eq}$ = 1000 K. This is close enough to the equilibrium temperature of TOI-5027 *b* (1056 K) to have negligible effects in its metal mass content estimate. For TOI-3837 *b*, this difference of 180 K in equilibrium temperature can entail a difference in core mass fraction of ΔCMF = 0.02. This small correction should be taken into account when comparing our estimate for TOI-3837 *b* with other interior models.

Table 5 shows the mean and uncertainties obtained in the interior structure retrievals. Given the degeneracy between CMF and envelope metallicity, which can only be solved by atmospheric characterization data, the interior models span a wide range of atmospheric metallicities. Nonetheless, our analysis allows us to determine the bulk metal mass fraction of the planet. TOI-3837 *b* has a metal content between that of Jupiter and Saturn, and TOI-6628 *b* is compatible with a Saturn-like composition. In contrast, TOI-5027 *b* and TOI-2328 *b* have bulk metal mass fractions between Saturn's ($Z_{planet}$ ∼ 0.20) and Neptune's ($Z_{planet}$ = 0.80 – 0.90) values (Miguel & Vazan 2023, and references therein). To compare the bulk metal content of these four planets with the extrasolar warm gas giants population, we show the mass-bulk metallicity trend in Fig. 9. The star metal content is computed with the metallicity of their respective host star, [Fe/H] (Table 1), as $Z_{star} = 0.0152 \times 10^{[Fe/H]}$. We calculate the uncertainties of the $Z_{planet}/Z_{star}$ ratio in Fig. 9 by sampling the MCMC's posterior distribution function of the planet's bulk metal mass fraction, by assuming a Gaussian distribution of the stellar [Fe/H] using the bootstrap method. TOI-2328 *b* and TOI-3837 *b* follow closely the exoplanet mass-bulk metallicity trend, indicating the these two planets likely formed via core accretion (Thorngren et al. 2016). TOI-6628 *b*'s metal content ratio is compatible within uncertainties with Jupiter. In contrast, TOI-5027 *b* is slightly more metal-rich than the 1$\sigma$ estimate of the exoplanet trend. This indicates that TOI-5027 *b* may have formed by accreting both pebbles and vapour enriched gas during the runaway accretion phase (Helled & Morbidelli 2021; Bitsch & Mah 2023).

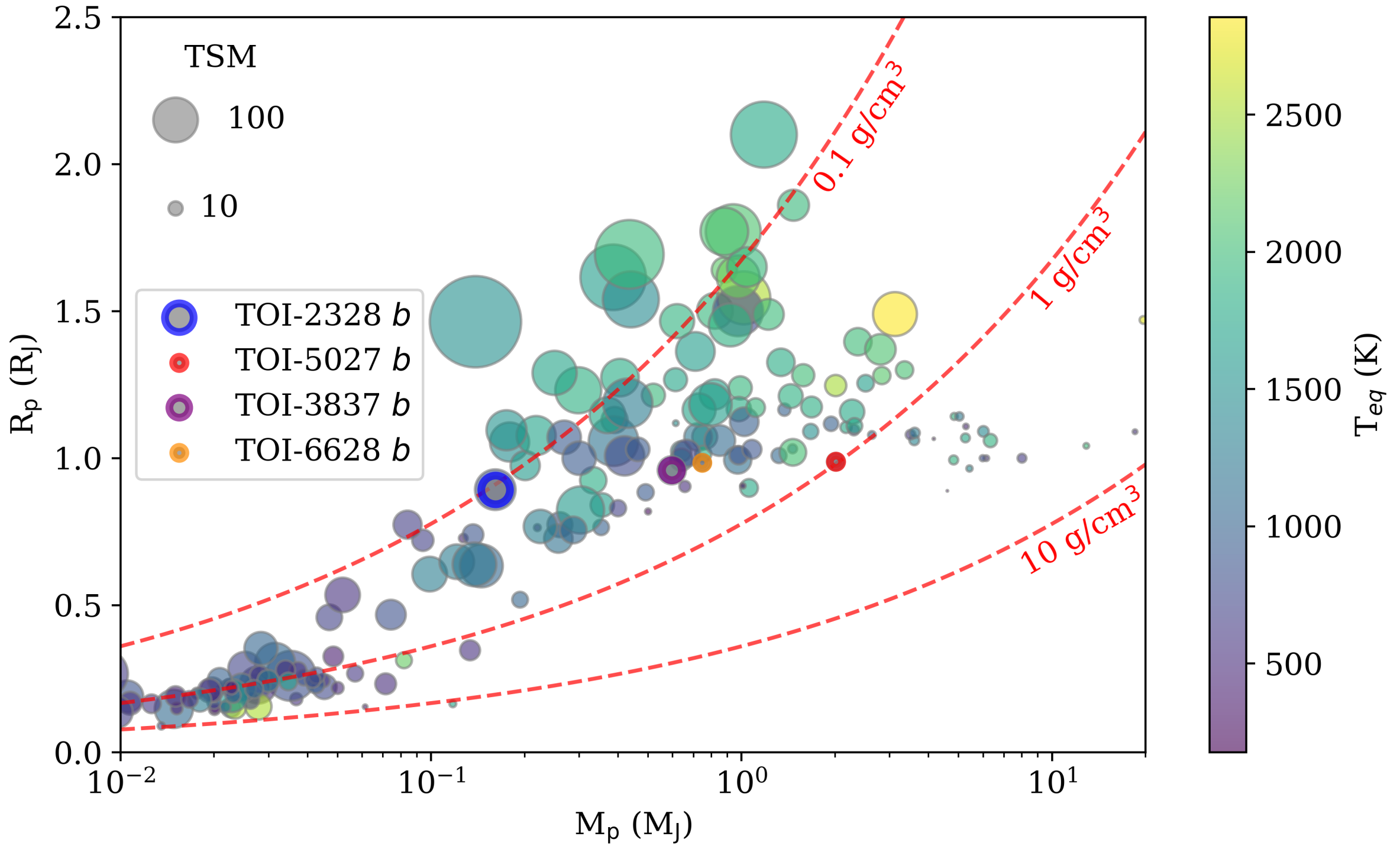

**Fig. 11.** Planet radius as a function of the planet mass for the population of warm giant planets (P > 10 days), color coded by equilibrium temperature. The size of the circles scales with the transit spectroscopy metric (TSM; Kempton et al. (2018)). Dashed red lines correspond to bulk densities of 0.1, 1, and 10 g cm$^{-3}$.

For comparison, we also computed interior models for the three gas giants presented here (TOI-6628 *b*, TOI-3837 *b* and TOI-5027 *b*), using the modules for experiments in stellar astrophysics (MESA, Paxton et al. 2011). For this, we mainly followed the implementation presented in Jones et al. (2024), but here we computed models with all of the heavy elements in the core, and a pure H/He gaseous envelope. In addition, here we estimated the density of the core using the equation of state presented in Hubbard & Marley (1989), using a 7:3 mixture (in mass) of rock and ice. The resulting models that better reproduce the position of these planets in the age-radius diagram are presented in Figure 10, and the corresponding planet metallicity and heavy-element enrichment with respect to the host star are listed in Table 5. The results obtained by these two sets of models agree well within 1-$\sigma$.

## 6. Discussion

TOI-6628 *b*, TOI-3837 *b*, TOI-5027 *b* and TOI-2328 *b* are four giant plants with masses between 0.1 and 2 Jupiter masses, orbital periods between 10 and 20 days and predicted time-averaged equilibrium temperatures below 1200 K. Figure 11 show the population of well-characterized transiting exoplanets in the period-radius plane, size-coded by the transit spectroscopy metric (TSM, Kempton et al. (2018)) and color-coded by planetary mass. About 1/3 of the planets have periods longer than 10 days, and all four planets here presented belong to that group.

Given their distance to their host star, irradiation levels in the planet surface from the host star should be mild enough to neglect proximity effects in the study of the planet's internal bulk structure. Their time-averaged surface temperatures are all below 1200 K and their incident flux is below $2\times10^8$ erg/s. Because of this, warm Jupiters have, in general, radii similar to or below 1 $R_J$, while hot Jupiters have radii well above the predictions according to classical planetary structural models (Zapolsky & Salpeter 1969).

Three of our targets orbit their host star in eccentric orbits, the exception being TOI-2328 *b*. We highlight the case of TOI-6628 *b*, which has a measured eccentricity of 0.67, and the case of TOI-5027 *b*, which has a measured eccentricity of 0.39. They are among the few planets orbiting in high eccentricity orbits. In the case of TOI-3837 *b*, we measured an eccentricity of 0.19, however the Δlog-evidence between the eccentric and non-eccentric model is less than 5, suggesting that the current dataset could be also explained by a circular orbit. Winn

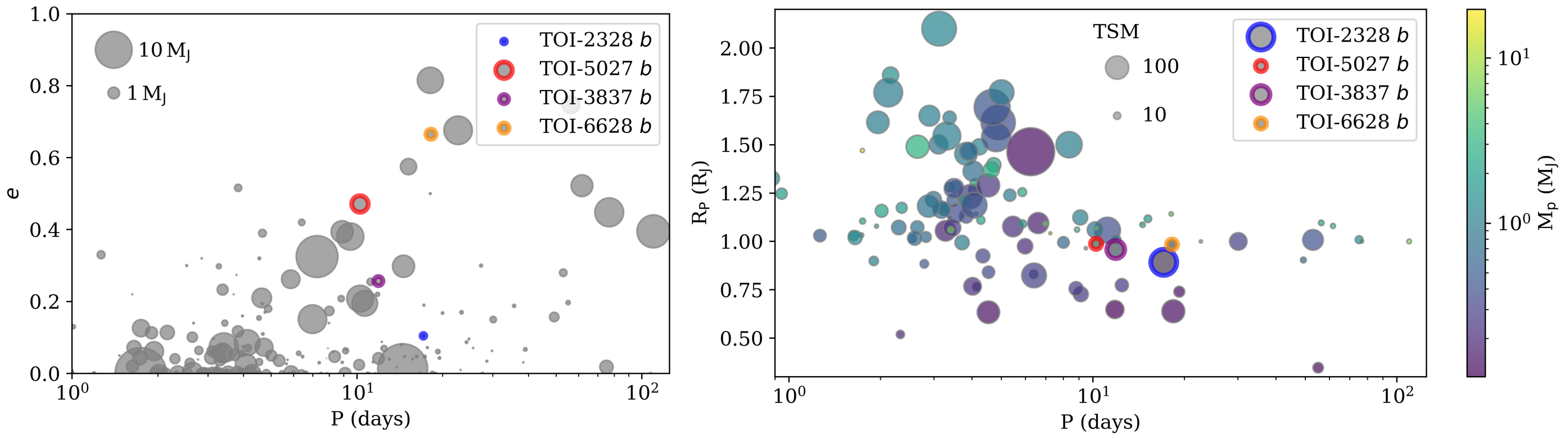

**Fig. 12.** *Left panel*: orbital eccentricity as a function of orbital period for the population of transiting giant planets with a period of up to 100 days. TOI-6028 *b* and TOI-5027 *b* are among the few discovered high eccentricity planets. *Right panel*: Planet radius vs. orbital period diagram for the population of giant planets, color-coded by planet mass. The size of the circles scales with the TSM. The four systems presented in this study belong to the population of long period planets, with P > 10 days.

& Fabrycky (2015) predicted a wide distribution of eccentricities for Warm Giants, as opposed to the circular orbits of HJs. Our findings provide additional evidence to such theoretical expectations. Figure 12 show the population of well-characterized transiting exoplanets in the period-eccentricity plane, size-coded by planetary mass. All four planets lie in the region of this plane that is still sparsely populated. Only 5% of the total population of giant planets detected so far have eccentricities larger than 0.5, and TOI-6628 *b* is one of them. Moreover, about 2/3 of the total population of transiting exoplanets have circular orbits with $e < 0.1$. Hence, TOI-6628 *b*, TOI-3837 *b* and TOI-5027 *b* belong to an emerging population of still rare warm giants in rather eccentric orbits.

The long-term trends in the RV residuals of TOI-6628 *b*, TOI-3837 *b* and TOI-5027 *b* suggest the presence of possible outer companions. However, it is not possible to constrain the properties of the outer companion with the current dataset. A long-term RV follow-up campaign should provide better constraints to the observed linear trends.

Orbital obliquities measured using the Rossiter-McLaughlin effect (R-M) can be a powerful tool to constrain warm Jupiter migration theories (Petrovich & Tremaine 2016). By assuming an impact parameter *b* of zero, the upper limits for the amplitudes of the R-M effect for TOI-6628 *b*, TOI-3837 *b*, TOI-5027 *b* and TOI-2328 *b* are 31.3 m/s, 23.4 m/s, 54.5 m/s and 27.1 m/s, respectively. This makes this set of planets well-suited targets for the measurement of the projected angle between the stellar and orbital angular momenta. By studying the obliquity distribution of the growing population of warm jupiters, we will be able to further understand giant planets formation and evolution.

Planet interior models suggest that, most likely, TOI-6628 *b*, TOI-3837 *b*, TOI-5027 *b* and TOI-2328 *b* internal structure consists of an icy and rocky core and a gaseous envelope composed mainly by hydrogen and helium, providing supporting evidence to the core accretion theory of planet formation. While MESA models consider that the metals are concentrated in the core, GASTLI consider that the metals are mixed in the envelope. MESA models incorporate the effect of stellar evolution in the planet interior structure, which can play an important role in the case of stars that are evolving off the Main Sequence, increasing their luminosity, as it seems to be the case of TOI-3837. Both GASTLI and MESA models are consistsent and in excellent agreement within the uncertainties.

## 7. Conclusions

We report the discovery of planetary companions orbiting TOI-6628, TOI-3837, TOI-5027 and TOI-2328. TOI-6628 *b* has a mass of 0.75 $M_J$ and orbits its host star every 18.18 days in an eccentric orbit with $e = 0.67$. TOI-3837 *b* has a mass of 0.59 $M_J$ and an eccentricity of 0.19. TOI-5027 *b* is the more massive of the four with a mass of 2.01 $M_J$ and has an eccentricity of 0.39. TOI-2328 *b* has a mass of 0.16 $M_J$ and orbits its host star in circular orbit.

By building up a sample of fully characterized Warm Jupiters, it will be possible to study the orbital parameters distributions and provide further constrains to planetary formation and evolution models.

*Acknowledgements.* M. T. P. acknowledges the support of Fondecyt-ANID fellowship no. 3210253 and ASTRON-0037. A.J. R.B. and V.S. acknowledge support from ANID – Millennium Science Initiative – ICN12_009 and AIM23-0001. R.B. acknowledges support from FONDECYT Project 1241963. A.J. acknowledges support from FONDECYT project 1210718.

| Parameter | Distribution | Value |
|---|---|---|
| P (days) | $\mathcal{U}$(18.17, 18.19) | $18.184236\pm^{1.27e-05}_{1.32e-05}$ |
| $t_0$ (days) | $\mathcal{U}$(2458602.71, 2458602.73) | $2458602.7209\pm^{8.59e-04}_{8.58e-04}$ |
| b | $\mathcal{U}$(0, 1) | $0.81\pm^{1.58e-02}_{1.71e-02}$ |
| p | $\mathcal{U}$(0, 1) | $0.1\pm^{1.24e-03}_{1.09e-03}$ |
| $e$ | $\mathcal{U}$(0, 1) | $0.67\pm^{1.58e-02}_{1.59e-02}$ |
| $\omega$ (deg) | $\mathcal{U}$(0, 360) | $215.41\pm^{2.42}_{2.32}$ |
| $\rho$ | $\mathcal{N}(\rho, \sigma_\rho)$ | $1299.31\pm^{8.93e+01}_{9.5e+01}$ |
| K (m/s) | $\mathcal{U}$(0, 1000) | $78.9\pm^{3.10}_{3.12}$ |
| $\mu_{\rm HARPS}$ (m/s) | $\mathcal{U}$(-1000, 1000) | $34.89\pm^{3.40}_{3.54}$ |
| $\mu_{\rm FEROS}$ (m/s) | $\mathcal{U}$(-1000, 1000.) | $-19.56\pm^{5.59}_{5.79}$ |
| $\sigma_{\rm HARPS}$ (m/s) | $\mathcal{U}$(0.1, 100.) | $5.42\pm^{3.45}_{4.76}$ |
| $\sigma_{\rm FEROS}$ (m/s) | $\mathcal{U}$(0.1, 100.) | $5.02\pm^{4.13}_{4.63}$ |
| $q_{1,\rm TESS}$ | $\mathcal{U}$(0, 1) | $0.22\pm^{1.12e-01}_{9.82e-02}$ |
| $q_{2,\rm TESS}$ | $\mathcal{U}$(0, 1) | $0.31\pm^{2.63e-01}_{2.04e-01}$ |
| $q_{1,\rm OMES-CDK600}$ | $\mathcal{U}$(0, 1) | $0.39\pm^{1.17e-01}_{1.02e-01}$ |
| $q_{2,\rm OMES-CDK600}$ | $\mathcal{U}$(0, 1) | $0.75\pm^{1.65e-01}_{1.89e-01}$ |
| $q_{1,\rm Hazelwood}$ | $\mathcal{U}$(0, 1) | $0.47\pm^{2.43e-01}_{2.37e-01}$ |
| $q_{2,\rm Hazelwood}$ | $\mathcal{U}$(0, 1) | $0.55\pm^{2.79e-01}_{2.92e-01}$ |
| $\mu_{\rm TESS11}$ | $\mathcal{N}$(0, 0.1) | $0.0\pm^{2.07e-05}_{2.20e-05}$ |
| $\mu_{\rm TESS38}$ | $\mathcal{N}$(0, 0.1) | $-0.0\pm^{1.67e-05}_{1.69e-05}$ |
| $\mu_{\rm TESS65}$ | $\mathcal{N}$(0, 0.1) | $0.0\pm^{1.22e-05}_{1.22e-05}$ |
| $\mu_{\rm OMES-CDK600}$ | $\mathcal{N}$(0, 0.1) | $-0.0\pm^{1.47e-04}_{1.53e-04}$ |
| $\mu_{\rm Hazelwood}$ | $\mathcal{N}$(0, 0.1) | $0.0\pm^{5.78e-03}_{5.75e-03}$ |
| $\sigma_{\rm TESS11}$ | $\mathcal{J}$(0.1, 1000) | $3.71\pm^{5.04e+01}_{3.34}$ |
| $\sigma_{\rm TESS38}$ | $\mathcal{J}$(0.1, 1000) | $3.36\pm^{3.22e+01}_{3.05}$ |
| $\sigma_{\rm TESS65}$ | $\mathcal{J}$(0.1, 1000) | $999.98\pm^{1.75e-02}_{3.34e-02}$ |
| $\sigma_{\rm OMES-CDK600}$ | $\mathcal{J}$(0.1, 1000) | $956.66\pm^{3.15e+01}_{5.31e+01}$ |
| $\sigma_{\rm Hazelwood}$ | $\mathcal{J}$(0.1, 1000) | $10.0\pm^{2.29e+02}_{9.47}$ |
| $\theta_{0,\rm Hazelwood}$ | $\mathcal{U}$(-100, 100) | $0.0\pm^{4.95e-03}_{5.09e-03}$ |
| $m_{d,\rm TESS11}$ | fixed, 1.0 | – |
| $m_{d,\rm TESS38}$ | fixed, 1.0 | – |
| $m_{d,\rm TESS65}$ | fixed, 1.0 | – |
| $m_{d,\rm OMES-CDK600}$ | fixed, 1.0 | – |
| $m_{d,\rm Hazelwood}$ | fixed, 1.0 | – |

**Table 6.** Prior Parameter Distributions for the Joint Transit and Radial Velocity Analysis of TOI-6628 *b*.

| Parameter | Distribution | Value |
|---|---|---|
| P (days) | $\mathcal{U}$(11.88, 11.90) | $11.88865\pm^{2.86e-05}_{2.68e-05}$ |
| $t_0$ (days) | $\mathcal{U}$(2458876.27, 2458876.29) | $2459530.161\pm^{7.39e-04}_{7.98e-04}$ |
| b | $\mathcal{U}$(0, 1) | $0.74\pm^{3.6e-02}_{2.5e-02}$ |
| p | $\mathcal{U}$(0, 1) | $0.08\pm^{1.57e-03}_{1.79e-03}$ |
| $e$ | $\mathcal{U}$(0, 1) | $0.22\pm^{4.59e-02}_{5.80e-02}$ |
| $\omega$ (deg) | $\mathcal{U}$(0, 360) | $280.56\pm^{1.24e+01}_{1.17e+01}$ |
| $\rho$ | $\mathcal{N}(\rho, \sigma_\rho)$ | $736.05\pm^{3.81e+01}_{4.01e+01}$ |
| K (m/s) | $\mathcal{U}$(0, 1000) | $57.72\pm^{3.97}_{3.98}$ |
| $\mu_{\rm HARPS}$ (m/s) | $\mathcal{U}$(-1000, 1000.) | $16.28\pm^{8.49}_{8.58}$ |
| $\mu_{\rm FEROS}$ (m/s) | $\mathcal{U}$(-1000, 1000.) | $-6.83\pm^{6.80}_{6.34}$ |
| $\sigma_{\rm HARPS}$ (m/s) | $\mathcal{U}$(0.1, 100) | $3.53\pm^{3.74}_{3.64}$ |
| $\sigma_{\rm FEROS}$ (m/s) | $\mathcal{U}$(0.1, 100) | $7.57\pm^{4.77}_{4.49}$ |
| $q_{1,\rm TESS}$ | $\mathcal{U}$(0, 1) | $0.24\pm^{1.38e-01}_{8.38e-02}$ |
| $q_{2,\rm TESS}$ | $\mathcal{U}$(0, 1) | $0.62\pm^{2.52e-01}_{3.34e-01}$ |
| $q_{1,\rm KeplerCam}$ | $\mathcal{U}$(0, 1) | $0.79\pm^{1.45e-01}_{1.92e-01}$ |
| $q_{2,\rm KeplerCam}$ | $\mathcal{U}$(0, 1) | $0.27\pm^{2.94e-01}_{1.90e-01}$ |
| $\mu_{\rm TESS21}$ | $\mathcal{N}$(0, 0.1) | $-0.0\pm^{1.96e-05}_{1.86e-05}$ |
| $\mu_{\rm TESS45}$ | $\mathcal{N}$(0, 0.1) | $-0.0\pm^{1.65e-05}_{1.72e-05}$ |
| $\mu_{\rm TESS46}$ | $\mathcal{N}$(0, 0.1) | $-0.0\pm^{1.62e-05}_{1.53e-05}$ |
| $\mu_{\rm TESS48}$ | $\mathcal{N}$(0, 0.1) | $0.0\pm^{1.68e-05}_{1.75e-05}$ |
| $\mu_{\rm KeplerCam}$ | $\mathcal{N}$(0, 0.1) | $0.0\pm^{2.07e-03}_{1.95e-03}$ |
| $\sigma_{\rm TESS21}$ | $\mathcal{U}$(0.1, 1000) | $1.93\pm^{1.67}_{1.68}$ |
| $\sigma_{\rm TESS45}$ | $\mathcal{U}$(0.1, 1000) | $3.89\pm^{2.33}_{3.52}$ |
| $\sigma_{\rm TESS46}$ | $\mathcal{U}$(0.1, 1000) | $4.53\pm^{2.73}_{4.07}$ |
| $\sigma_{\rm TESS48}$ | $\mathcal{U}$(0.1, 1000) | $8.05\pm^{8.45_{7.57}}$ |
| $\sigma_{\rm KeplerCam}$ | $\mathcal{U}$(0.1, 1000) | $984.04\pm^{1.18e+01}_{2.43e+01}$ |
| $m_{d,\rm TESS21}$ | fixed, 1.0 | – |
| $m_{d,\rm TESS45}$ | fixed, 1.0 | – |
| $m_{d,\rm TESS46}$ | fixed, 1.0 | – |
| $m_{d,\rm TESS48}$ | fixed, 1.0 | – |
| $m_{d,\rm KeplerCam}$ | fixed, 1.0 | – |

**Table 7.** Prior Parameter Distributions for the Joint Transit and Radial Velocity Analysis of TOI-3837 *b*.

| Parameter | Distribution | Value |
|---|---|---|
| P (days) | $\mathcal{U}$(10.23, 10.25) | $10.243678\pm^{8.82e-06}_{8.84e-06}$ |
| $t_0$ (days) | $\mathcal{U}$(2458649.44, 2458649.46) | $2458649.458\pm^{9.79e-04}_{9.4e-04}$ |
| b | $\mathcal{U}$(0, 1) | $0.93\pm^{2.61e-02}_{1.75e-02}$ |
| p | $\mathcal{U}$(0, 1) | $0.11\pm^{1.29e-02}_{6.42e-03}$ |
| $e$ | $\mathcal{U}$(0, 1) | $0.4\pm^{3.25e-02}_{2.86e-02}$ |
| $\omega$ (deg) | $\mathcal{U}$(0, 360) | $288.56\pm^{3.37}_{3.37}$ |
| $\rho$ | $\mathcal{N}(\rho, \sigma_\rho)$ | $1833.27\pm^{3.88e+01}_{3.75e+01}$ |
| K (m/s) | $\mathcal{U}$(0, 1000) | $201.77\pm^{4.00}_{4.23}$ |
| $\mu_{\rm FEROS}$ (m/s) | $\mathcal{U}$(-1000, 1000.) | $6.15\pm^{7.15}_{7.45}$ |
| $\sigma_{\rm FEROS}$ (m/s) | $\mathcal{U}$(0.1, 100.) | $28.32\pm^{3.48}_{3.63}$ |
| $q_{1,\rm TESS}$ | $\mathcal{U}$(0, 1) | $0.19\pm^{2.40e-01}_{1.34e-01}$ |
| $q_{2,\rm TESS}$ | $\mathcal{U}$(0.1, 100.) | $0.27\pm^{2.90e-01}_{1.84e-01}$ |
| $q_{1,\rm LCOGT}$ | $\mathcal{U}$(0, 1) | $0.36\pm^{1.93e-01}_{1.43e-01}$ |
| $q_{2,\rm LCOGT}$ | $\mathcal{U}$(0, 1) | $0.51\pm^{3.05e-01}_{3.17e-01}$ |
| $q_{1,\rm OMES-RiDK500}$ | $\mathcal{U}$(0, 1) | $0.37\pm^{2.65e-01}_{1.95e-01}$ |
| $q_{2,\rm OMES-RiDK500}$ | $\mathcal{U}$(0, 1) | $0.38\pm^{3.33e-01}_{2.72e-01}$ |
| $\mu_{\rm TESS12}$ | $\mathcal{N}$(0, 0.1) | $0.0\pm^{1.85e-05}_{1.74e-05}$ |
| $\mu_{\rm TESS39}$ | $\mathcal{N}$(0, 0.1) | $0.0\pm^{1.43e-05}_{1.43e-05}$ |
| $\mu_{\rm TESS66}$ | $\mathcal{N}$(0, 0.1) | $-0.0\pm^{1.59e-05}_{1.49e-05}$ |
| $\mu_{\rm LCOGT}$ | $\mathcal{N}$(0, 0.1) | $-0.0\pm^{1.12e-04}_{1.12e-04}$ |
| $\mu_{\rm OMES-RiDK500}$ | $\mathcal{N}$(0, 0.1) | $-0.0\pm^{2.68e-04}_{2.46e-04}$ |
| $\sigma_{\rm TESS12}$ | $\mathcal{J}$(0.1, 1000) | $185.99\pm^{3.75e+01}_{5.28e+01}$ |
| $\sigma_{\rm TESS39}$ | $\mathcal{J}$(0.1, 1000) | $288.54\pm^{3.55e+01}_{3.71e+01}$ |
| $\sigma_{\rm TESS66}$ | $\mathcal{J}$(0.1, 1000) | $385.66\pm^{5.65e+01}_{6.76e+01}$ |
| $\sigma_{\rm LCOGT}$ | $\mathcal{J}$(0.1, 1000) | $670.81\pm^{5.77e+01}_{5.46e+01}$ |
| $\sigma_{\rm OMES-RiDK500}$ | $\mathcal{J}$(0.1, 1000) | $10.9\pm^{2.76e+02}_{1.03e+01}$ |
| $m_{d,\rm TESS12}$ | fixed, 1.0 | – |
| $m_{d,\rm TESS39}$ | fixed, 1.0 | – |
| $m_{d,\rm TESS66}$ | fixed, 1.0 | – |
| $m_{d,\rm LCOGT}$ | fixed, 1.0 | – |
| $m_{d,\rm OMES-RiDK500}$ | fixed, 1.0 | – |

**Table 8.** Prior Parameter Distributions for the Joint Transit and Radial Velocity Analysis of TOI-5027 *b*.

| Parameter | Distribution | Value |
|---|---|---|
| P (days) | $\mathcal{U}$(17.09, 17.11) | $17.101966\pm^{3.84e-06}_{5.46e-06}$ |
| $t_0$ (days) | $\mathcal{U}$(2458330.47, 2458330.49) | $2458330.4894\pm^{3.51e-04}_{2.55e-04}$ |
| b | $\mathcal{U}$(0, 1) | $0.69\pm^{2.16e-02}_{1.81e-02}$ |
| p | $\mathcal{U}$(0, 1) | $0.1\pm^{3.34e-04}_{3.e-04}$ |
| $e$ | $\mathcal{U}$(0, 1) | $0.06\pm^{4.6e-02}_{2.86e-02}$ |
| $\omega$ (deg) | $\mathcal{U}$(0, 360) | $148.75\pm^{1.41e+01}_{9.31}$ |
| $\rho$ | $\mathcal{N}(\rho, \sigma_\rho)$ | $1806.12\pm^{4.21e+01}_{3.54e+01}$ |
| K (m/s) | $\mathcal{U}$(0, 1000) | $13.29\pm^{1.57}_{1.53}$ |
| $\mu_{\rm HARPS}$ (m/s) | $\mathcal{U}$(-1000, 1000) | $-6.81\pm^{6.04}_{4.95}$ |
| $\mu_{\rm FEROS}$ (m/s) | $\mathcal{U}$(-1000, 1000) | $-2.96\pm^{4.33}_{3.84}$ |
| $\mu_{\rm PFS}$ (m/s) | $\mathcal{U}$(-1000, 1000) | $-11.17\pm^{3.7}_{5.73}$ |
| $\sigma_{\rm HARPS}$ (m/s) | $\mathcal{U}$(0.1, 100) | $20.76\pm^{4.52}_{5.13}$ |
| $\sigma_{\rm FEROS}$ (m/s) | $\mathcal{U}$(0.1, 100) | $24.89\pm^{1.49}_{2.23}$ |
| $\sigma_{\rm PFS}$ (m/s) | $\mathcal{U}$(0.1, 100) | $21.72\pm^{6.11}_{1.11e+01}$ |
| $q_{1,\rm TESS}$ | $\mathcal{U}$(0, 1) | $0.03\pm^{1.41e-02}_{1.18e-02}$ |
| $q_{2,\rm TESS}$ | $\mathcal{U}$(0, 1) | $0.84\pm^{9.67e-02}_{8.69e-02}$ |
| $q_{1,\rm OMES-CDK600}$ | $\mathcal{U}$(0, 1) | $0.45\pm^{3.24e-02}_{3.71e-02}$ |
| $q_{2,\rm OMES-CDK600}$ | $\mathcal{U}$(0, 1) | $0.8\pm^{7.27e-02}_{7.25e-02}$ |
| $q_{1,\rm CHAT}$ | $\mathcal{U}$(0, 1) | $0.39\pm^{3.59e-02}_{4.26e-02}$ |
| $q_{2,\rm CHAT}$ | $\mathcal{U}$(0, 1) | $0.83\pm^{5.91e-02}_{7.32e-02}$ |
| $q_{1,\rm Hazelwood}$ | $\mathcal{U}$(0, 1) | $0.68\pm^{8.16e-02}_{8.23e-02}$ |
| $q_{2,\rm Hazelwood}$ | $\mathcal{U}$(0, 1) | $0.78\pm^{8.16e-02}_{1.43e-01}$ |
| $q_{1,\rm LCOGT}$ | $\mathcal{U}$(0, 1) | $0.05\pm^{3.34e-02}_{2.27e-02}$ |
| $q_{2,\rm LCOGT}$ | $\mathcal{U}$(0, 1) | $0.6\pm^{1.22e-01}_{2.38e-01}$ |
| $q_{1,\rm LCOGT-g}$ | $\mathcal{U}$(0, 1) | $0.57\pm^{8.97e-02}_{9.37e-02}$ |
| $q_{2,\rm LCOGT-g}$ | $\mathcal{U}$(0, 1) | $0.26\pm^{1.44e-01}_{1.37e-01}$ |
| $\mu_{\rm TESS1}$ | $\mathcal{N}$(0, 0.1) | $-0.0\pm^{5.47e-06}_{6.55e-06}$ |
| $\mu_{\rm TESS13}$ | $\mathcal{N}$(0, 0.1) | $-0.0\pm^{1.67e-05}_{1.62e-05}$ |
| $\mu_{\rm TESS28}$ | $\mathcal{N}$(0, 0.1) | $-0.0\pm^{1.26e-05}_{1.06e-05}$ |
| $\mu_{\rm TESS39}$ | $\mathcal{N}$(0, 0.1) | $-0.0\pm^{6.31e-06}_{7.61e-06}$ |
| $\mu_{\rm TESS67}$ | $\mathcal{N}$(0, 0.1) | $-0.0\pm^{8.32e-06}_{6.17e-06}$ |
| $\mu_{\rm TESS68}$ | $\mathcal{N}$(0, 0.1) | $-0.0\pm^{1.23e-05}_{9.68e-06}$ |
| $\mu_{\rm OMES-CDK600}$ | $\mathcal{N}$(0, 0.1) | $0.0\pm^{5.53e-05}_{1.24e-04}$ |
| $\mu_{\rm CHAT}$ | $\mathcal{N}$(0, 0.1) | $-0.0\pm^{8.51e-05}_{1.46e-04}$ |
| $\mu_{\rm Hazelwood}$ | $\mathcal{N}$(0, 0.1) | $0.01\pm^{4.41e-03}_{3.43e-03}$ |
| $\mu_{\rm LCOGT}$ | $\mathcal{N}$(0, 0.1) | $0.0\pm^{6.98e-05}_{6.66e-05}$ |
| $\mu_{\rm LCOGT-g}$ | $\mathcal{N}$(0, 0.1) | $0.0\pm^{1.01e-04}_{1.27e-04}$ |
| $\sigma_{\rm TESS1}$ | $\mathcal{J}$(0.1, 1000) | $342.6\pm^{1.10e+01}_{1.31e+01}$ |
| $\sigma_{\rm TESS13}$ | $\mathcal{J}$(0.1, 1000) | $859.98\pm^{2.42e+01}_{2.27e+01}$ |
| $\sigma_{\rm TESS28}$ | $\mathcal{J}$(0.1, 1000) | $34.3\pm^{4.33e+01}_{2.15e+01}$ |
| $\sigma_{\rm TESS39}$ | $\mathcal{J}$(0.1, 1000) | $9.63\pm^{2.17e+01}_{6.14}$ |
| $\sigma_{\rm TESS67}$ | $\mathcal{J}$(0.1, 1000) | $6.84\pm^{3.35e+01}_{5.18}$ |
| $\sigma_{\rm TESS68}$ | $\mathcal{J}$(0.1, 1000) | $8.43\pm^{2.27e+01}_{5.72}$ |
| $\sigma_{\rm OMES-CDK600}$ | $\mathcal{J}$(0.1, 1000) | $994.72\pm^{3.08}_{3.32}$ |
| $\sigma_{\rm CHAT}$ | $\mathcal{J}$(0.1, 1000) | $996.19\pm^{2.43}_{6.18}$ |
| $\sigma_{\rm Hazelwood}$ | $\mathcal{J}$(0.1, 1000) | $0.37\pm^{2.52}_{2.13e-01}$ |
| $\sigma_{\rm LCOGT}$ | $\mathcal{J}$(0.1, 1000) | $975.94\pm^{1.28e+01}_{1.73e+01}$ |
| $\sigma_{\rm LCOGT-g}$ | $\mathcal{J}$(0.1, 1000) | $807.0\pm^{1.33e+02}_{2.95e+02}$ |
| $m_{d,\rm TESS1}$ | fixed, 1.0 | – |
| $m_{d,\rm TESS13}$ | fixed, 1.0 | – |
| $m_{d,\rm TESS28}$ | fixed, 1.0 | – |
| $m_{d,\rm TESS39}$ | fixed, 1.0 | – |
| $m_{d,\rm TESS67}$ | fixed, 1.0 | – |
| $m_{d,\rm TESS68}$ | fixed, 1.0 | – |
| $m_{d,\rm OMES-CDK600}$ | fixed, 1.0 | – |
| $m_{d,\rm CHAT}$ | fixed, 1.0 | – |
| $m_{d,\rm Hazelwood}$ | fixed, 1.0 | – |
| $m_{d,\rm LCOGT}$ | fixed, 1.0 | – |
| $m_{d,\rm LCOGT-g}$ | fixed, 1.0 | – |

**Table 9.** Prior Parameter Distributions for the Joint Transit and Radial Velocity Analysis of TOI-2328 *b*.

## Appendix A: Spectroscopic data

**Table A.1.** RVs and stellar activity indices for TOI 6628.

| BJD (days) | RV (m/s) | BIS | Hα | $Ca_{II}$ | $He_I$ | $Na_{II}$ | Instrument |
|---|---|---|---|---|---|---|---|
| 2459268.84 | 7.9±8.3 | -0.036±0.013 | 0.125±0.003 | 0.135±0.015 | 0.518±0.007 | 0.189±0.004 | FEROS |
| 2459282.76 | 42.9±9.5 | 0.001±0.014 | 0.137±0.004 | 0.15±0.018 | 0.509±0.009 | 0.194±0.006 | FEROS |
| 2459291.877 | 18.6±9.5 | -0.013±0.013 | 0.116±0.003 | -0.218±0.046 | 0.512±0.006 | 0.213±0.004 | HARPS |
| 2459295.882 | -136.2±9.5 | -0.021±0.012 | 0.122±0.003 | -0.598±0.038 | 0.502±0.006 | 0.206±0.004 | HARPS |
| 2459417.588 | 8.0±10.8 | -0.017±0.015 | 0.109±0.004 | 0.167±0.029 | 0.523±0.01 | 0.177±0.006 | FEROS |
| 2459419.58 | 8.3±9.5 | -0.025±0.012 | 0.114±0.003 | -0.062±0.033 | 0.507±0.005 | 0.195±0.004 | HARPS |
| 2459421.634 | -50.0±20.3 | -0.046±0.027 | 0.113±0.005 | -0.345±0.102 | 0.506±0.009 | 0.184±0.008 | HARPS |
| 2459422.595 | 89.6±15.7 | -0.003±0.021 | 0.122±0.004 | -0.643±0.079 | 0.511±0.008 | 0.207±0.006 | HARPS |
| 2459423.609 | 7.7±15.6 | 0.017±0.021 | 0.109±0.004 | 0.356±0.081 | 0.524±0.007 | 0.221±0.006 | HARPS |
| 2459424.572 | -13.7±10.4 | -0.021±0.014 | 0.114±0.003 | -0.145±0.042 | 0.517±0.006 | 0.194±0.004 | HARPS |
| 2459425.579 | -30.6±13.9 | -0.016±0.018 | 0.117±0.004 | -0.364±0.06 | 0.525±0.007 | 0.19±0.005 | HARPS |
| 2459449.552 | -93.8±17.7 | -0.045±0.023 | 0.12±0.004 | -0.163±0.09 | 0.495±0.008 | 0.199±0.007 | HARPS |
| 2459657.867 | 18.7±10.4 | -0.004±0.014 | 0.127±0.003 | -0.325±0.035 | 0.497±0.006 | 0.243±0.004 | HARPS |
| 2459658.747 | 3.7±8.1 | 0.003±0.011 | 0.129±0.002 | -0.308±0.02 | 0.498±0.005 | 0.227±0.003 | HARPS |
| 2459666.816 | -18.7±7.5 | -0.028±0.01 | 0.149±0.002 | -0.151±0.016 | 0.505±0.005 | 0.222±0.003 | HARPS |
| 2459669.832 | -130.6±17.2 | -0.031±0.022 | 0.136±0.004 | -18.728±5.437 | 0.49±0.008 | 0.289±0.007 | HARPS |
| 2459711.621 | -91.7±8.1 | -0.015±0.011 | 0.126±0.002 | -0.337±0.018 | 0.499±0.005 | 0.268±0.004 | HARPS |
| 2459713.634 | -10.3±15.3 | -0.048±0.02 | 0.129±0.004 | -1.9±0.095 | 0.478±0.008 | 0.327±0.007 | HARPS |
| 2459736.768 | -79.1±14.0 | -0.009±0.018 | 0.15±0.004 | -4.589±0.308 | 0.458±0.007 | 0.326±0.006 | HARPS |
| 2459750.664 | -119.4±15.6 | 0.01±0.02 | 0.137±0.004 | -24.954±7.034 | 0.482±0.008 | 0.25±0.007 | HARPS |
| 2459829.516 | 24.0±28.1 | -0.087±0.037 | 0.154±0.007 | 2.709±0.126 | 0.422±0.012 | 0.297±0.012 | HARPS |
| 2459834.498 | 18.5±17.7 | -0.014±0.023 | 0.149±0.004 | -10.363±1.137 | 0.475±0.008 | 0.25±0.007 | HARPS |
| 2459838.484 | -44.1±14.0 | -0.023±0.018 | 0.136±0.004 | -3.708±0.296 | 0.487±0.007 | 0.244±0.006 | HARPS |

**Table A.2.** RVs and activity indices for TOI 3837.

| BJD (days) | RV (m/s) | BIS | Hα | $Ca_{II}$ | $He_I$ | $Na_{II}$ | Instrument |
|---|---|---|---|---|---|---|---|
| 2459213.835 | 59.3±13.6 | 0.064±0.014 | 0.142±0.004 | 0.11±0.018 | 0.506±0.01 | 0.39±0.008 | FEROS |
| 2459219.811 | -36.4±10.3 | -0.037±0.012 | 0.124±0.003 | 0.125±0.013 | 0.498±0.007 | 0.365±0.006 | FEROS |
| 2459223.851 | 26.2±10.5 | -0.005±0.012 | 0.116±0.003 | 0.145±0.02 | 0.492±0.007 | 0.378±0.006 | FEROS |
| 2459249.786 | -59.8±7.6 | 0.008±0.009 | 0.102±0.002 | -0.084±0.015 | 0.508±0.005 | 0.401±0.004 | HARPS |
| 2459251.764 | -45.5±7.0 | -0.016±0.008 | 0.108±0.002 | 0.005±0.012 | 0.506±0.005 | 0.399±0.004 | HARPS |
| 2459255.731 | 16.6±7.5 | -0.011±0.009 | 0.108±0.002 | -0.044±0.012 | 0.502±0.005 | 0.398±0.004 | HARPS |
| 2459260.673 | -47.8±8.2 | -0.001±0.009 | 0.11±0.002 | -0.061±0.014 | 0.504±0.005 | 0.393±0.004 | HARPS |
| 2459261.744 | 37.9±12.5 | 0.008±0.015 | 0.107±0.004 | -0.183±0.026 | 0.507±0.007 | 0.407±0.007 | HARPS |
| 2459264.671 | -59.2±7.6 | 0.007±0.009 | 0.109±0.002 | -0.068±0.013 | 0.499±0.005 | 0.402±0.004 | HARPS |
| 2459265.713 | -15.9±9.7 | 0.003±0.011 | 0.116±0.003 | 0.1±0.014 | 0.501±0.006 | 0.367±0.005 | FEROS |
| 2459274.686 | -15.8±12.4 | -0.004±0.013 | 0.126±0.004 | 0.086±0.013 | 0.503±0.009 | 0.385±0.008 | FEROS |
| 2459275.715 | 40.7±12.4 | -0.007±0.013 | 0.107±0.003 | 0.105±0.026 | 0.496±0.008 | 0.381±0.006 | FEROS |
| 2459277.663 | 60.9±11.9 | 0.022±0.013 | 0.12±0.004 | 0.079±0.015 | 0.49±0.009 | 0.38±0.007 | FEROS |
| 2459281.659 | -65.9±11.9 | -0.003±0.013 | 0.119±0.003 | 0.124±0.021 | 0.503±0.008 | 0.358±0.007 | FEROS |
| 2459294.719 | 47.0±18.3 | -0.009±0.022 | 0.126±0.005 | -1.26±0.076 | 0.517±0.009 | 0.422±0.008 | HARPS |

**Table A.3.** RVs and activity indices of TOI 5027.

| BJD (days) | RV (m/s) | BIS | Hα | $Ca_{II}$ | $He_I$ | $Na_{II}$ | Instrument |
|---|---|---|---|---|---|---|---|
| 2459683.815 | -31.1±13.1 | -0.009±0.013 | 0.155±0.004 | 0.135±0.011 | 0.482±0.008 | 0.353±0.007 | FEROS |
| 2459687.768 | -134.0±8.7 | 0.019±0.01 | 0.122±0.002 | 0.15±0.007 | 0.491±0.005 | 0.387±0.004 | FEROS |
| 2459690.843 | -1.2±10.1 | -0.016±0.011 | 0.122±0.002 | 0.199±0.01 | 0.495±0.006 | 0.387±0.005 | FEROS |
| 2459759.8 | -180.6±14.0 | -0.032±0.014 | -0.046±0.002 | 0.281±0.02 | 0.526±0.009 | 0.35±0.007 | FEROS |
| 2459765.769 | 208.0±12.7 | 0.005±0.013 | 0.027±0.003 | 0.153±0.019 | 0.491±0.008 | 0.354±0.006 | FEROS |
| 2459767.71 | 47.9±11.5 | 0.061±0.012 | 0.328±0.005 | 0.121±0.012 | 0.488±0.007 | 0.382±0.006 | FEROS |
| 2459773.648 | -135.4±10.7 | -0.005±0.011 | 0.14±0.002 | 0.202±0.01 | 0.479±0.006 | 0.383±0.005 | FEROS |
| 2459835.538 | 221.7±12.4 | -0.019±0.013 | -0.111±0.004 | 0.144±0.014 | 0.499±0.008 | 0.403±0.006 | FEROS |
| 2459845.59 | -25.7±10.7 | 0.002±0.011 | 0.106±0.002 | 0.278±0.05 | 0.498±0.005 | 0.389±0.004 | FEROS |
| 2459847.583 | 55.9±15.2 | 0.008±0.015 | 0.137±0.004 | 0.061±0.023 | 0.523±0.01 | 0.421±0.009 | FEROS |
| 2459850.599 | -201.2±10.9 | 0.021±0.011 | 0.128±0.002 | 0.12±0.031 | 0.507±0.006 | 0.358±0.005 | FEROS |
| 2459857.491 | 217.4±15.5 | 0.074±0.015 | 0.158±0.004 | 0.286±0.022 | 0.486±0.01 | 0.378±0.008 | FEROS |
| 2459859.515 | 82.3±10.3 | 0.055±0.011 | 0.122±0.003 | 0.168±0.012 | 0.485±0.006 | 0.392±0.005 | FEROS |
| 2459861.537 | -56.2±12.8 | 0.023±0.013 | 0.156±0.003 | 0.276±0.021 | 0.502±0.007 | 0.379±0.006 | FEROS |
| 2459868.518 | -120.3±8.8 | 0.005±0.01 | 0.112±0.002 | 0.132±0.035 | 0.487±0.005 | 0.376±0.004 | FEROS |
| 2459871.503 | 94.9±12.6 | 0.054±0.013 | 0.132±0.003 | 0.215±0.022 | 0.492±0.008 | 0.353±0.006 | FEROS |
| 2459872.496 | 168.8±18.0 | 0.014±0.017 | 0.139±0.005 | 0.336±0.039 | 0.519±0.011 | 0.386±0.01 | FEROS |
| 2460030.874 | -109.0±8.7 | 0.004±0.01 | 0.121±0.002 | 0.156±0.004 | 0.495±0.005 | 0.378±0.004 | FEROS |
| 2460031.845 | -54.6±8.6 | -0.011±0.01 | 0.127±0.002 | 0.125±0.005 | 0.501±0.005 | 0.379±0.004 | FEROS |
| 2460033.865 | -95.0±9.3 | -0.049±0.013 | 0.167±0.003 | 0.344±0.029 | 0.514±0.008 | 0.113±0.004 | FEROS |
| 2460034.826 | -71.9±7.9 | 0.032±0.009 | 0.07±0.001 | 0.15±0.005 | 0.505±0.004 | 0.382±0.003 | FEROS |
| 2460047.892 | 109.8±10.9 | 0.029±0.011 | 0.009±0.002 | 0.111±0.009 | 0.511±0.006 | 0.206±0.004 | FEROS |
| 2460063.474 | 68.1±14.8 | -0.141±0.018 | 0.112±0.004 | 0.371±0.152 | 0.481±0.012 | 0.271±0.009 | FEROS |

**Table A.4.** RVs and activity indices of TOI-2328.

| BJD (days) | RV (m/s) | BIS | H$\alpha$ | $Ca_{II}$ | $He_{I}$ | $Na_{II}$ | Instrument |
|---|---|---|---|---|---|---|---|
| 2458670.851 | 16.1±18.5 | 0.18±0.024 | 0.171±0.008 | 0.5±0.375 | 0.479±0.016 | 0.339±0.015 | FEROS |
| 2458674.86 | -82.1±14.5 | -0.029±0.019 | 0.138±0.005 | 0.36±0.152 | 0.519±0.012 | 0.381±0.011 | FEROS |
| 2458675.819 | -105.1±9.5 | 0.006±0.014 | 0.128±0.003 | 0.165±0.046 | 0.481±0.008 | 0.257±0.006 | FEROS |
| 2458718.802 | -5.0±7.7 | -0.012±0.012 | 0.124±0.003 | 0.243±0.026 | 0.493±0.006 | 0.247±0.005 | FEROS |
| 2458721.819 | 17.7±8.2 | -0.024±0.012 | 0.124±0.003 | 0.128±0.036 | 0.497±0.007 | 0.224±0.004 | FEROS |
| 2458722.69 | 44.5±7.8 | 0.003±0.012 | 0.125±0.003 | 0.304±0.034 | 0.488±0.007 | 0.231±0.004 | FEROS |
| 2458723.819 | 11.2±12.1 | 0.022±0.017 | 0.136±0.005 | 0.179±0.054 | 0.502±0.012 | 0.275±0.009 | FEROS |
| 2458724.828 | 17.5±8.9 | -0.027±0.013 | 0.136±0.003 | 0.142±0.027 | 0.498±0.008 | 0.235±0.006 | FEROS |
| 2458725.71 | -22.8±9.6 | 0.026±0.014 | 0.119±0.003 | 0.271±0.043 | 0.508±0.009 | 0.254±0.006 | FEROS |
| 2458740.689 | 38.0±12.4 | -0.263±0.016 | 0.133±0.004 | 0.229±0.038 | 0.48±0.01 | 0.344±0.009 | FEROS |
| 2458741.767 | 37.7±8.9 | -0.085±0.013 | 0.121±0.003 | 0.136±0.026 | 0.496±0.007 | 0.258±0.005 | FEROS |
| 2458785.637 | -1.8±10.1 | -0.018±0.015 | 0.138±0.004 | 0.065±0.031 | 0.5±0.009 | 0.269±0.007 | FEROS |
| 2458787.688 | 16.3±11.7 | -0.037±0.016 | 0.138±0.005 | 0.476±0.063 | 0.503±0.011 | 0.296±0.009 | FEROS |
| 2458794.614 | 21.0±8.3 | -0.033±0.012 | 0.115±0.003 | 0.168±0.027 | 0.498±0.007 | 0.24±0.005 | FEROS |
| 2458796.699 | 95.0±8.6 | -0.065±0.012 | 0.119±0.002 | 0.086±0.07 | 0.491±0.006 | 0.263±0.004 | FEROS |
| 2458797.627 | -39.5±7.8 | -0.07±0.012 | 0.122±0.003 | 0.212±0.027 | 0.501±0.006 | 0.264±0.005 | FEROS |
| 2458799.636 | 45.6±10.6 | -0.103±0.014 | 0.139±0.004 | 0.378±0.042 | 0.493±0.008 | 0.277±0.006 | FEROS |
| 2458801.75 | 23.7±12.2 | -0.087±0.016 | 0.133±0.005 | 0.119±0.04 | 0.5±0.011 | 0.29±0.008 | FEROS |
| 2458803.673 | -231.0±8.3 | -0.014±0.012 | 0.125±0.003 | 0.161±0.025 | 0.505±0.007 | 0.235±0.005 | FEROS |
| 2458805.656 | -35.8±8.3 | -0.011±0.012 | 0.131±0.003 | 0.088±0.04 | 0.51±0.007 | 0.242±0.005 | FEROS |
| 2458808.71 | 53.3±8.9 | -0.006±0.013 | 0.127±0.003 | 0.106±0.033 | 0.508±0.008 | 0.241±0.006 | FEROS |
| 2458810.715 | -54.5±9.8 | 0.007±0.014 | 0.138±0.004 | 0.207±0.035 | 0.502±0.009 | 0.254±0.006 | FEROS |
| 2458831.609 | -11.7±6.4 | -0.016±0.008 | 0.118±0.002 | 0.105±0.017 | 0.494±0.004 | 0.237±0.003 | HARPS |
| 2458834.604 | -13.5±10.3 | -0.02±0.013 | 0.119±0.003 | 0.002±0.03 | 0.506±0.006 | 0.246±0.005 | HARPS |
| 2458835.651 | -9.4±7.4 | -0.011±0.01 | 0.112±0.002 | 0.052±0.021 | 0.496±0.004 | 0.242±0.003 | HARPS |
| 2458836.62 | 5.0±6.0 | -0.016±0.008 | 0.112±0.002 | 0.11±0.018 | 0.504±0.004 | 0.246±0.003 | HARPS |
| 2458840.562 | -15.3±6.0 | -0.012±0.008 | 0.113±0.002 | 0.093±0.017 | 0.502±0.004 | 0.231±0.003 | HARPS |
| 2458850.543 | 26.5±8.0 | -0.016±0.01 | 0.121±0.003 | 0.067±0.018 | 0.49±0.005 | 0.246±0.004 | HARPS |
| 2458853.596 | -2.2±8.7 | -0.011±0.011 | 0.115±0.003 | 0.136±0.027 | 0.501±0.005 | 0.228±0.004 | HARPS |
| 2458874.542 | -0.2±12.4 | -0.015±0.016 | 0.117±0.003 | 0.039±0.033 | 0.49±0.006 | 0.231±0.005 | HARPS |
| 2458882.539 | 9.7±9.4 | 0.005±0.012 | 0.117±0.002 | 0.101±0.024 | 0.507±0.005 | 0.231±0.004 | HARPS |
| 2458888.533 | 12.1±10.3 | 0.019±0.013 | 0.114±0.003 | 0.033±0.027 | 0.494±0.006 | 0.243±0.004 | HARPS |
| 2458894.548 | 13.7±9.4 | -0.006±0.012 | 0.112±0.002 | -0.049±0.03 | 0.499±0.005 | 0.219±0.004 | HARPS |
| 2458899.525 | -17.7±10.3 | -0.017±0.013 | 0.113±0.003 | 0.022±0.036 | 0.509±0.006 | 0.238±0.004 | HARPS |
| 2459177.625 | -10.4±9.4 | -0.006±0.012 | 0.122±0.003 | -0.214±0.038 | 0.496±0.006 | 0.285±0.005 | HARPS |
| 2459179.624 | 29.6±9.4 | -0.009±0.012 | 0.122±0.003 | -0.324±0.033 | 0.5±0.006 | 0.288±0.005 | HARPS |
| 2459182.619 | -5.9±8.7 | -0.018±0.011 | 0.124±0.003 | -0.35±0.03 | 0.496±0.006 | 0.264±0.004 | HARPS |
| 2459192.604 | 11.1±5.2 | -0.015±0.007 | 0.117±0.002 | -0.114±0.015 | 0.494±0.004 | 0.251±0.003 | HARPS |
| 2459204.563 | -9.2±5.9 | -0.007±0.008 | 0.12±0.002 | 0.112±0.016 | 0.487±0.004 | 0.254±0.003 | HARPS |
| 2459211.628 | 14.3±8.0 | -0.006±0.01 | 0.126±0.003 | -0.192±0.027 | 0.49±0.006 | 0.258±0.004 | HARPS |
| 2459421.894 | 72.7±10.2 | 0.002±0.013 | 0.111±0.003 | 0.009±0.028 | 0.507±0.006 | 0.306±0.005 | HARPS |
| 2459422.886 | 8.0±11.3 | -0.029±0.015 | 0.106±0.003 | -0.02±0.031 | 0.503±0.006 | 0.298±0.005 | HARPS |
| 2459423.881 | 0.5±13.8 | -0.031±0.018 | 0.111±0.004 | -0.023±0.046 | 0.496±0.007 | 0.3±0.006 | HARPS |
| 2459425.926 | -21.4±8.0 | -0.023±0.01 | 0.112±0.002 | 0.07±0.02 | 0.496±0.004 | 0.3±0.004 | HARPS |
| 2459452.819 | 6.0±2.3 | — | — | — | — | — | PFS |
| 2459471.84 | 15.9±2.2 | — | — | — | — | — | PFS |
| 2459473.818 | 13.6±3.0 | — | — | — | — | — | PFS |
| 2459505.748 | 14.9±2.5 | — | — | — | — | — | PFS |
| 2459506.742 | 6.4±2.1 | — | — | — | — | — | PFS |
| 2459534.697 | -2.5±2.2 | — | — | — | — | — | PFS |
| 2459854.732 | -8.4±2.2 | — | — | — | — | — | PFS |
| 2459854.746 | -10.8±1.9 | — | — | — | — | — | PFS |
| 2459865.752 | 15.1±1.9 | — | — | — | — | — | PFS |
| 2459865.765 | 20.4±2.0 | — | — | — | — | — | PFS |
| 2459895.683 | -7.2±2.7 | — | — | — | — | — | PFS |
| 2459895.701 | -1.3±2.5 | — | — | — | — | — | PFS |

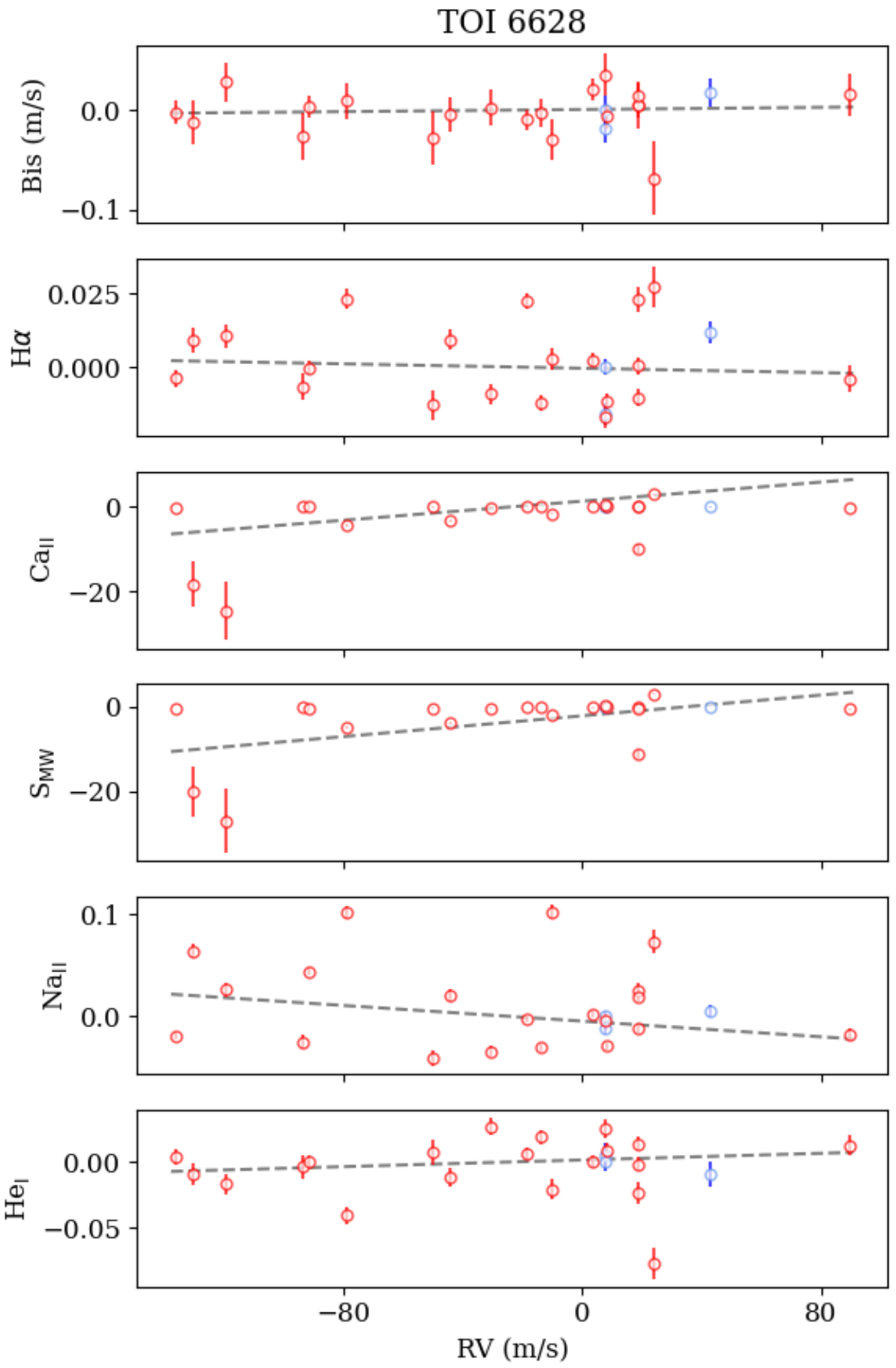

Fig. B.1. Stellar activity indices plotted as a function of the RV for TOI 6628. The gray dashed line show the best linear fit to the data.

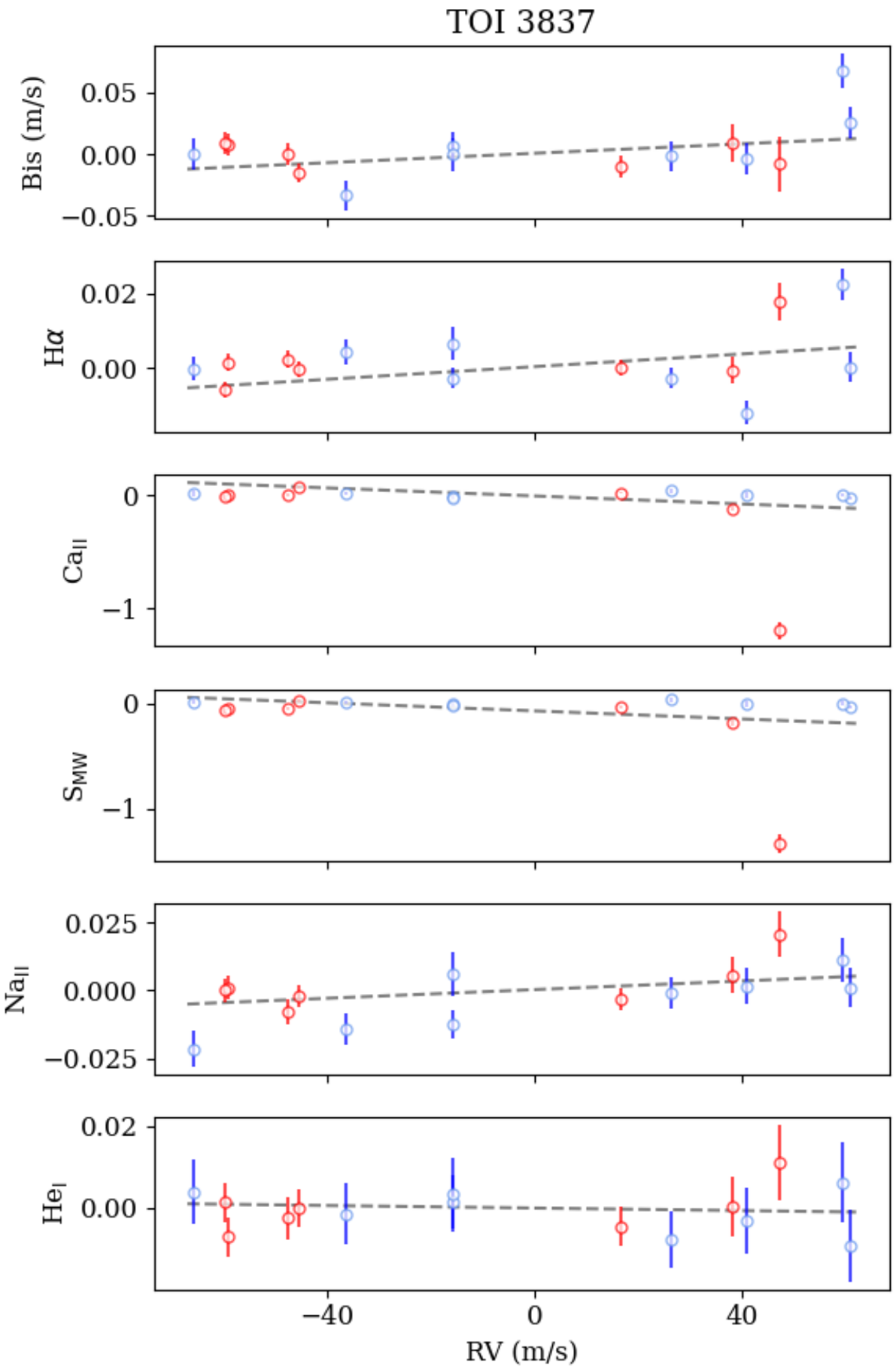

Fig. B.2. Stellar activity indices plotted as a function of the RV for TOI 3837. The gray dashed line show the best linear fit to the data.

## Appendix B: Activity indices

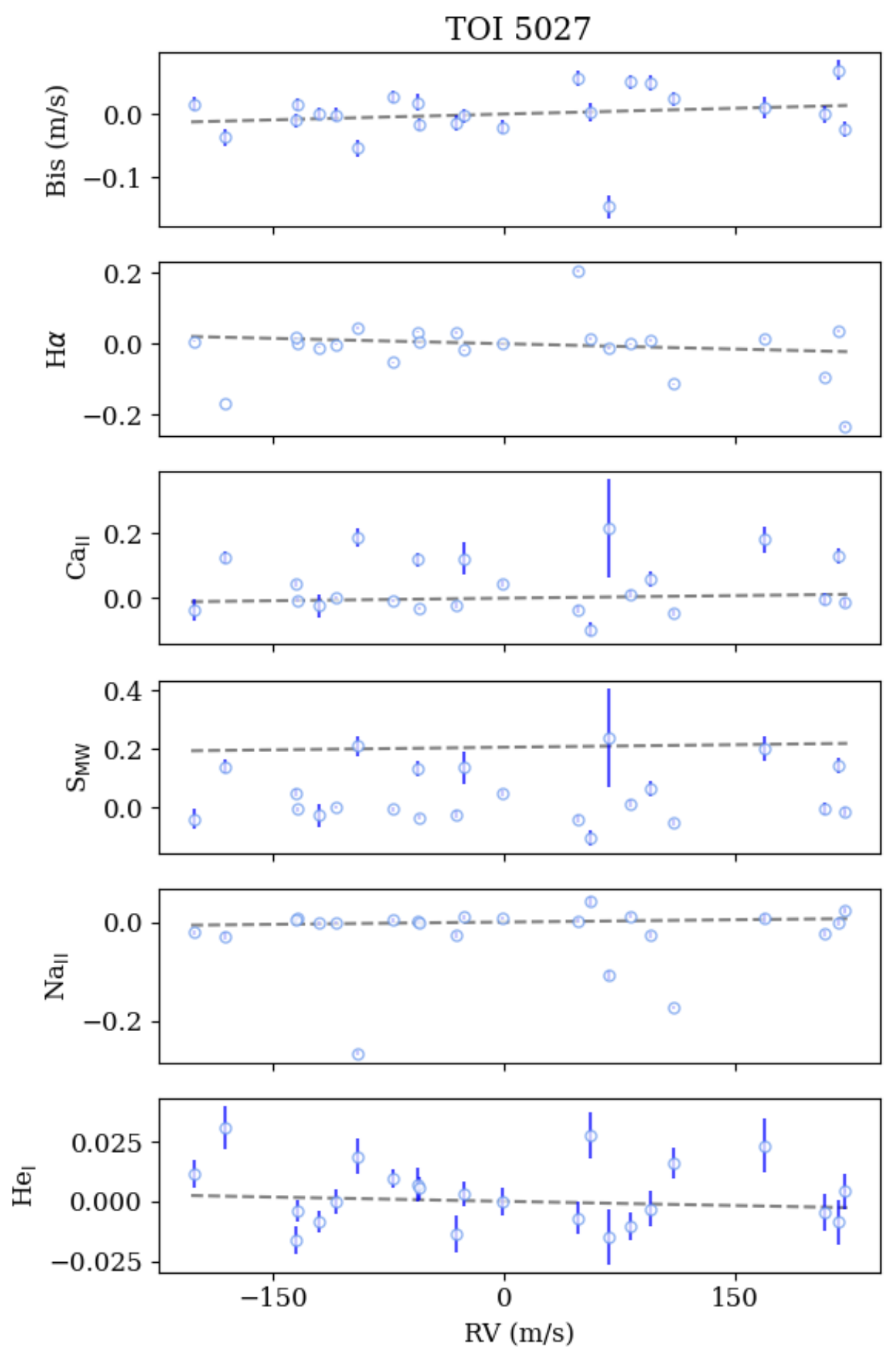

**Fig. B.3.** Stellar activity indices plotted as a function of the RV for TOI 5027. The gray dashed line show the best linear fit to the data.

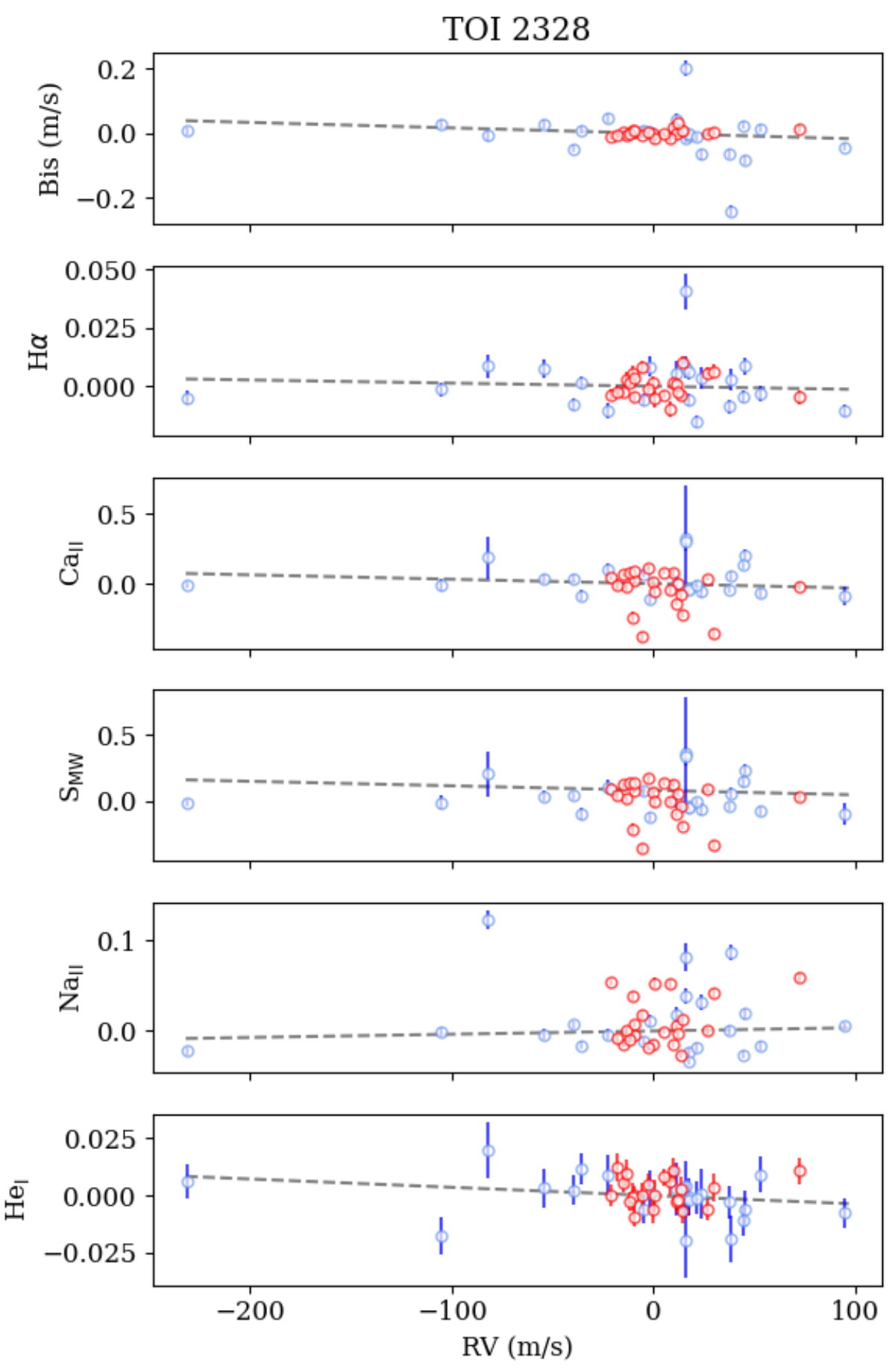

**Fig. B.4.** Stellar activity indices plotted as a function of the RV for TOI 2328. The gray dashed line show the best linear fit to the data.